\documentclass[iop,twocolappendix]{emulateapj}
\usepackage{amsmath,mathtools,mathrsfs}
\newcommand{\sgra}{Sgr~A$^*$}

\begin{document}
\title{A Parametric model for the shapes of black-hole shadows in non-Kerr spacetimes}
\author{Lia Medeiros\altaffilmark{1,2,$\dagger$, $\star$}, Dimitrios Psaltis\altaffilmark{2}, Feryal \"Ozel\altaffilmark{2}}

\altaffiltext{1}{School of Natural Sciences, Institute for Advanced Study, 1 Einstein Drive, Princeton, NJ 08540}
\altaffiltext{2}{Steward Observatory and Department of Astronomy, University of Arizona, 933 N. Cherry Ave., Tucson, AZ 85721}

\altaffiltext{$\dagger$}{NSF Astronomy and Astrophysics Postdoctoral Fellow}
\altaffiltext{$\star$}{lia@ias.edu}

\begin{abstract}
The Event Horizon Telescope (EHT) is taking the first images of black holes resolved at horizon scales to measure their shadows and probe accretion physics. A promising avenue for testing the hypothesis that astrophysical black holes are described by the Kerr solution to Einstein's equations is to compare the size and shape of the shadow a black hole casts on the surrounding emission to the predictions of the Kerr metric. We develop here an efficient parametric framework to perform this test. We carry out ray-tracing simulations for several parametrized non-Kerr metrics to create a large data set of non-Kerr shadows that probe the allowed parameter space for the free parameters of each metric. We then perform principal components analysis (PCA) on this set of shadows and show that only a small number of components are needed to accurately reconstruct all shadows within the set. We further show that the amplitude of the PCA components are smoothly related to the free parameters in the metrics and, therefore, that these PCA components can be fit to EHT observations in order to place constraints on the free parameters of these metrics that will help quantify any potential deviations from the Kerr solution. 
\end{abstract}

\section{Introduction}\label{sec:intro}
Much of our current understanding of black holes relies on the assumption that they are described by the Kerr solution to the Einstein equations. One possible avenue for conducting a test of this assumption with observations in the electromagnetic spectrum is by measuring the size and shape of the shadow a black hole casts on the surrounding emission~\citep{2010ApJ...718..446J,Psaltis2018}. 

\begin{figure*}[t!]
\centering
\includegraphics{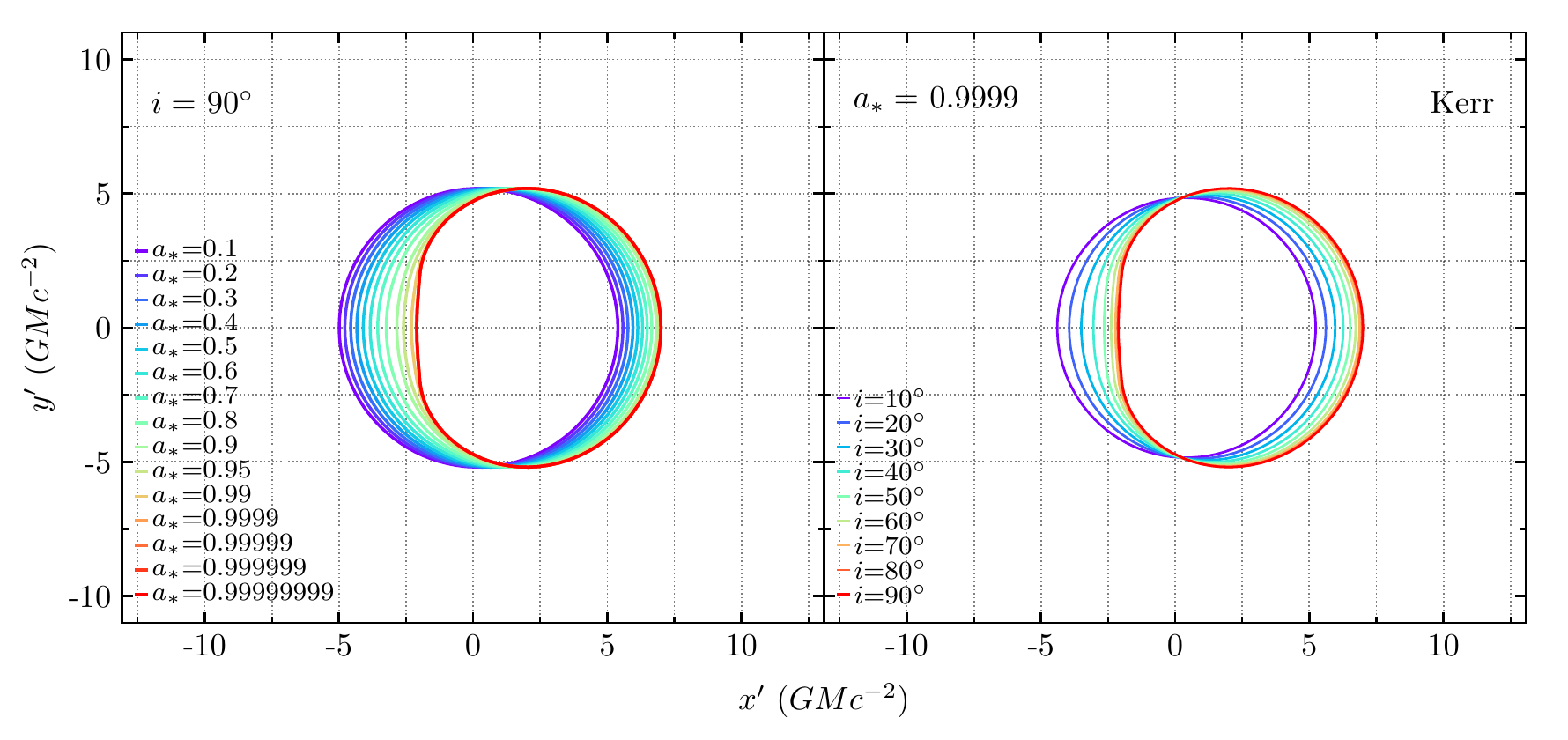}
\caption{Several black hole shadows that result from the Kerr metric as a function of the black hole spin, $a_*$ (right), and the inclination of the observer, $i$ (left). In this and all of the following figures, we assume that the spin vector points upwards. For all but the most rapidly spinning black holes viewed by equatorial observers, the shadow has a nearly circular shape.}
\label{fig:params_Kerr}
\end{figure*}

The black hole shadow depends only on the geometry of the spacetime and not on the astrophysics of the accretion process. For a Kerr black hole, the shadow has a radius of $5\pm 0.2 GMc^{-2}$ for all spins and observer inclinations (see e.g., \citealt{2000CQGra..17..123D,2010ApJ...718..446J,  2013ApJ...777...13C}, see also Figure~\ref{fig:params_Kerr}). Therefore, measuring the size of the shadow of a black hole of known mass constitutes a null hypothesis test of the Kerr metric~\citep{2015ApJ...814..115P}. At the same time, the shape of the shadow for a Kerr black hole is nearly circular for all but the highest spins. For a general spacetime, the shape of the shadow depends primarily on the deviation of the spacetime quadrupole from its Kerr value~\citep{2010ApJ...718..446J}. As a result, measuring the shadow shape leads to a test of the General Relativistic no-hair theorem~\citep{Psaltis2016}.

The Event Horizon Telescope (EHT) is a mm-VLBI experiment that has produced the first ever image of the black hole in M87 resolved at horizon scales and has measured the size and shape of its shadow~\citep{EHTPaperI}. The initial analysis of the EHT data has revealed no striking deviations from the predictions for the Kerr metric~\citep{EHTPaperVI}; improved measurements with more interferometric stations and new algorithms will tighten these constraints in the near future. The second primary target of the EHT is the Galactic Center black hole, \sgra. This has the largest angular size on the sky of any currently known black hole  (e.g., \citealt{2012ApJ...758...30J})  and well constrained mass and distance \citep{2008ApJ...689.1044G, 2009ApJ...692.1075G,2018A&A...615L..15G}; therefore, measuring the shape and size of its shadow is expected to provide a precise test of the Kerr metric. 

In order to use the shape of the black hole shadow to test the Kerr metric with both Sgr~ A* and the black hole in M87, we need to explore the shapes and sizes of shadows in other metrics. Significant progress has been made in simulating the observational appearance of black holes and more exotic compact objects with different spacetimes and in different theories of gravity (see, e.g., \citealt{2009PhRvD..79d3002B,2010PhRvD..81l4045A,2010CQGra..27t5006B,2014PhRvD..90b4073P,2018NatAs...2..585M, 2019MNRAS.482...52S}). Additionally, several parametrized metrics have been developed that can be used to explore the range of non-Kerr spacetimes in a way that is agnostic to the underlying physical theory (for a few axisymmetric examples, see \citealt{1992CQGra...9.2477M,2006CQGra..23.4167G, 2010PhRvD..81b4030V, 2011PhRvD..83j4027V,2011PhRvD..83l4015J, 2016PhRvD..93f4015K}). The black hole shadows that result from some of these parameterized metrics have also been explored (see e.g., \citealt{2010ApJ...718..446J, 2013ApJ...777..170J, 2015EPJC...75..315G, 2016PhRvD..94h4025Y, 2017JCAP...10..051W}, and \citealt{Cunha2018} for a review).

Deriving a shadow from a given metric is relatively straightforward. However, using a shadow to place constraints on the metrics that could have created it presents a significant challenge as the mapping between the two is highly non-linear. \citet{2015MNRAS.454.2423A} proposed an algorithm that represents the shadow shapes empirically as a sum of Legendre polynomials, with no reference to the underlying metrics. Here, we use Principal Component Analysis to show that the shapes of all black-hole shadows generated by several of the parameterized metrics mentioned above can be represented by a small set of functions. This allows us to generate a general parametric model for shadow shapes that is largely agnostic of the underlying spacetime metric, is more compact than a general polynomial expansion, and can be used to derive metric constraints from the EHT data.

The paper is organized as follows. In \S 2, we introduce the parameterized metrics we consider; we outline our ray-tracing simulations in \S 3. In \S 4, we introduce principal components analysis and apply it to the simulated set of shadows. In \S 5, we assess the accuracy of the PCA reconstructions and explore the relationship between the amplitude of PCA components and the metric parameters. Finally we discuss the implications of our work in \S 6.

\section{Parametrically perturbed metrics}\label{sec:metrics}

In this section, we introduce a number of parameterized metrics and write them explicitly in the form we use in the simulations. We select a few metrics from those that have been published, while prioritizing metrics that differ significantly from each other. For example, since the metric of \citet{2011PhRvD..83j4027V} builds upon the metric of \citet{2004PhRvD..69l4022C} and the metric of \citet{2014PhRvD..89f4007C} builds upon the metric of \citet{2011PhRvD..83l4015J}, we choose to only include one metric from each of these pairs. 
We note that many of the proposed parameterized metrics have pathologies such as naked singularities, closed time-like loops, or non-Lorentzian geometries, that can significantly complicate our numerical calculations. Resolving such pathologies is outside the scope of the present work, so we restrict ourselves to three representative metrics that have been investigated in detail for pathologies in \citet{2013ApJ...777..170J} (in particular see Table 1 in this reference for a summary of such pathologies). 

Our starting point is the Kerr metric, which in Boyer-Lindquist coordinates takes the form (see, e.g., \citealt{1972ApJ...178..347B})
\begin{equation}
\begin{split}
g_{tt}^{\text{K}} &= - \left(1-\frac{2M r}{\Sigma}\right),\\
g_{rr}^{\text{K}} &= \frac{\Sigma}{\Delta} ,\,\,\,\,\,\,\,g_{\theta \theta}^{\text{K}} = \Sigma,\\
g_{\phi \phi}^{\text{K}} &= \left( r^2 + a^2 + \frac{2 M a^2 r \sin ^2 \theta}{\Sigma}\right)\sin^2 \theta,\\
g_{t \phi}^{\text{K}} &= - \frac{2 M a r \sin^2 \theta}{\Sigma},\\
\end{split}
\end{equation}
where 
\begin{equation}
\Delta \equiv a^2-2 M r +r^2, \,\,\,\,\,\,\,
\Sigma \equiv r^2 + a^2 \cos ^2\theta,
\end{equation}
$M$ is the black hole mass, and $a$ is the dimensional spin parameter, $a\equiv J/M$. We will also make use of the dimensionless spin parameter $a_*=J/M^2$. Here and throughout the paper, we use gravitational units and set $G=c=1$. In Figure~\ref{fig:params_Kerr}, we show, as a point of comparison to the following figures, the effect of changing the black hole spin and the inclination of the observer on the black hole shadow of the Kerr metric. As discussed earlier, the size and shape of a black hole shadow depends very weakly on spin; Kerr shadows are approximately circular, except for extremely high values of spin.

\subsection{The Quasi Kerr Metric }
The Quasi-Kerr (QK) metric is based on the work of Hartle and Thorne \citep{1967ApJ...150.1005H,1968ApJ...153..807H} for slowly spinning neutron stars and was adapted to describe general vacuum spacetimes by~\citet{2006CQGra..23.4167G}. This metric has all but the quadrupole moments equal to those of the Kerr metric; the quadrupole moment is set to
\begin{equation}
Q_{\mathrm{QK}} = -M (a^2 + \epsilon_{\mathrm{QK}}M^2),
\end{equation}
where $\epsilon_{\mathrm{QK}}$ measures the strength of the deviation from the Kerr quadrupole.
The metric is perturbed in contravariant form such that
\begin{equation}\label{eq:perturbcontra}
g^{\mu \nu}_{\mathrm{QK}} = g^{\mu\nu}_{\text{K}} +  \epsilon_{\mathrm{QK}} h^{\mu\nu}_{\mathrm{QK}} ,
\end{equation}
where $g^{\mu\nu}_{\text{K}}$ are the metric components for the Kerr metric in contravariant form and $h^{\mu\nu}_{\mathrm{QK}} $ are the metric perturbations for the QK metric. In Boyer-Lindquist-like coordinates, the elements of the perturbed metric are given by \citep{2006CQGra..23.4167G, 2013PhRvD..87l4017J}
\begin{equation}\label{eq:QKmetric}
\begin{split}
h^{tt}_{\mathrm{QK}} &= \frac{r}{r-2 M}(1-3\cos^2\theta )\mathcal{F}_1(r),\\
h^{rr}_{\mathrm{QK}}  &= \frac{r-2 M}{r}(1-3\cos^2\theta )\mathcal{F}_1(r),\\
h^{\theta\theta}_{\mathrm{QK}}  &= -\frac{1}{r^2}(1-3\cos^2\theta )\mathcal{F}_2(r),\\
h^{\phi\phi}_{\mathrm{QK}}  &= -\frac{1}{r^2\sin^2\theta}(1-3\cos^2\theta )\mathcal{F}_2(r),\\
\end{split}
\end{equation}
and
\begin{equation}\label{eq:F1F2}
\begin{split}
\mathcal{F}_1(r) &= -\frac{5 (r-M)}{8 M r (r-2M)} (2 M^2 + 6 M r - 3 r^2)\\
&-\frac{15r (r-2M)}{16 M^2} \ln
   \left(\frac{r}{r-2M}\right),\\
\mathcal{F}_2(r) &=  \frac{5}{8 M r} (2 M^2 - 3 M r - 3 r ^2) \\
& + \frac{15}{16M^2}(r^2 - 2 M^2)\ln
   \left(\frac{r}{r-2M}\right).\\
\end{split}
\end{equation}
The Quasi-Kerr metric is a solution to the vacuum Einstein equations for small spins, $a_*\ll 1$, and reduces to the Kerr metric when $\epsilon_{\mathrm{QK}}=0$. The validity of the Quasi-Kerr metric is limited to regions where $r\gtrsim 2M$ since the logarithm in eq. (\ref{eq:F1F2}) diverges at $r=2M$. In general, this metric describes a naked singularity \citep{2013ApJ...777..170J}. 

\citet{2010ApJ...716..187J} and \citet{2013PhRvD..87l4017J} found that, to first order in the deviation parameter, the conditions $a_*\lesssim 0.4$ and $-0.5 \lesssim \epsilon_{\mathrm{QK}} \lesssim 0.5$ are necessary but not sufficient to ensure that Lorentz violations and closed timelike loops are contained within the circular photon orbit. In this work, we solve the geodesics for the full metric shown in equations (\ref{eq:perturbcontra}) and (\ref{eq:QKmetric}) without any additional approximations or expansions. For this reason, the range of allowed values for $\epsilon_{\mathrm{QK}}$  is significantly smaller than the range explored in \citet{2010ApJ...718..446J}, who expanded the geodesic equations to first order in the perturbation parameter. In order to avoid pathologies in our work, we require, e.g., $\epsilon_{\mathrm{QK}}\lesssim 0.2$ for a black hole with $i=90^{\circ}$ and $a_*=0.4$ and $\epsilon_{\mathrm{QK}}\lesssim 0.35$ for a black hole with $i=90^{\circ}$ and $a_*=0.1.$\footnote{The ranges used for $\epsilon_{\mathrm{QK}}$ for various spins can be inferred from the length of the curves in Figure~\ref{fig:QK_x0_rscale}.}

\subsection{The Modified Kerr Metric}

This metric proposed by \citet{2011PhRvD..83l4015J} introduces polynomial perturbations to both the $rr-$ and $\theta\theta-$ components of the Schwarzchild metric and, following \citet{2011PhRvD..83j4027V}, uses the Newman-Janis algorithm (\citealt{NewmanJanis1965, 2000GReGr..32..445D}) to transform this into that of a rotating compact object. The resulting metric is not Ricci flat. Here we will follow the expansions of~\citet{2013ApJ...777..170J} and refer to this metric as JP (note that \citealt{2013ApJ...777..170J} refers to this metric as the modified Kerr metric or MK). 

In Boyer-Lindquist-like coordinates, the  components for the JP metric are
\begin{equation}
\begin{split}
g_{tt}^{\mathrm{JP}} &=-\frac{\tilde{\Sigma} \left(\Delta-a^2 A_2(r)^2 \sin ^2\theta
   \right)}{\left[A_1(r) \left(a^2+r^2\right)-a^2 A_2(r) \sin ^2\theta
   \right]^2},\\
g_{rr}^{\mathrm{JP}} &=\frac{\tilde{\Sigma}}{A_5(r) \Delta},\,\,\,\,\,\,\, g_{\theta \theta}^{\mathrm{JP}} =\tilde{\Sigma},\\
g_{\phi \phi}^{\mathrm{JP}} &=\frac{\tilde{\Sigma} \sin ^2\theta  \left[A_1(r)^2 \left(a^2+r^2\right)^2-a^2
   \Delta \sin ^2\theta \right]}{\left[A_1(r) \left(a^2+r^2\right)-a^2
   A_2(r) \sin ^2\theta \right]^2},\\
g_{t \phi}^{\mathrm{JP}} &=-\frac{a \tilde{\Sigma} \sin ^2\theta  \left[A_1(r) A_2(r)
   \left(a^2+r^2\right)-\Delta\right]}{\left[A_1(r) \left(a^2+r^2\right)-a^2
   A_2(r) \sin ^2\theta \right]^2},\\
\end{split}
\end{equation}
\\
where, to lowest order, 
\begin{equation}
\begin{split}
A_1(r) &\equiv\frac{\alpha_{13}M^3}{r^3}+1,\,\,\,\,\,\,\\ A_2(r) &\equiv \frac{\alpha_{22}M^2}{r^2}+1, \,\,\,\,\,\,\\
A_5(r) &\equiv \frac{\alpha_{52}M^2}{r^2}+1 ,\\
\tilde{\Sigma} &\equiv \Sigma + \frac{\epsilon_3 M^3}{r}.\\
\end{split}
\end{equation}
In this form, the metric has four free parameters, $\epsilon_3$, $\alpha_{13}$, $\alpha_{22}$, and $\alpha_{52}$. The Kerr metric is recovered when all free parameters are set to zero. \citet{2013ApJ...777..170J} showed that the shape of the black hole shadow only depends on the parameters $\alpha_{13}$ and $\alpha_{22}$. Therefore, here, we will only vary those two parameters and set $\epsilon_3$ and $\alpha_{52}$ to zero. 

This metric describes a rotating vacuum spacetime that deviates from the Kerr metric but does not have pathologies outside of the event horizon for all spins below unity as long as the perturbation parameters satisfy (see also Figure 2 in \citealt{2013ApJ...777..170J})
\begin{equation}
\begin{split}
\alpha_{13} &> \frac{-(M+\sqrt{M^2 - a^2})^3}{M^3},\\
\alpha_{22} &>\frac{-(M+\sqrt{M^2 - a^2})^2}{M^2}.\\
\end{split}
\end{equation}
For simplicity, in the present work we will only consider $ \alpha_{13}\ge-1.0$ and $ \alpha_{22}\ge-1.0$ since these are the lower limits that correspond to a maximally spinning black hole. Even though there is no upper limit on the values of these parameters, we simulate here shadows with $ \alpha_{13}\le 5$ and $ \alpha_{22}\le 5$. The event horizon for the JP metric does not depend on the perturbation parameters and coincides with the Kerr horizon,
\begin{equation}
r_{+} = M + \sqrt{M^2 - a^2},
\end{equation}
for all allowed values of the perturbation parameters.

\subsection{The Modified Gravity Bumpy Kerr Metric}

The metric proposed by \citet{2011PhRvD..83j4027V} uses bump functions to perturb the Kerr metric while still ensuring that the metric has three constants of motion. Here, we employ the metric in the form used by \citet{2011PhRvD..84f4016G} and \citet{2013PhRvD..87l4017J}, which makes some simplifications to the original metric to ensure certain properties (see \citealt{2011PhRvD..84f4016G} for details). We will employ the terminology in \citet{2013PhRvD..87l4017J} and refer to this metric as the modified gravity bumpy Kerr metric or MGBK. The MGBK metric is defined perturbatively in covariant form such that 
\begin{equation}
g_{\mu\nu}^{\mathrm{MGBK}} = g_{\mu\nu}^{\text{K}} + h_{\mu\nu}^{\mathrm{MGBK}},
\end{equation}
where the perturbations to each metric component are
\begin{widetext}
\begin{equation}
\begin{split}
h_{tt}^{\mathrm{MGBK}} &=-\frac{a  P_2}{M P_1} h_{t\phi}-\frac{a \Delta \Sigma^2}{2 M P_1} \frac{\partial{ h_{t\phi}}}{\partial{r}}-\frac{a^2 \Delta ^2  \Sigma  \sin ^2\theta}{2 M P_1} \frac{d \bar{\gamma}_1}{dr}-\frac{a \Delta ^2 \sin ^2\theta  \left(2 a^2 M r \sin ^2\theta
   +\hat{\Sigma}\right)}{2 M P_1} \frac{d \bar{\gamma}_3}{dr}\\
   &-\frac{a^2 \Delta ^2 \sin ^2\theta 
   (\Sigma -4 M r)}{2 M P_1} \frac{d \bar{\gamma}_4}{dr}+\frac{2 a^2 \Delta  \bar{\gamma}_1 r^2 \sin ^2\theta }{P_1}+\frac{\Delta 
  \bar{\gamma}_1 \hat{\rho}^2 \left(a^2+r^2\right)}{P_1}
   -\frac{a \Delta  \bar{\gamma}_3 P_3 \sin ^2\theta }{M \Sigma P_1}+\frac{2 \Delta  \bar{\gamma}_4 P_4}{\Sigma P_1} ,\\
h_{rr}^{\mathrm{MGBK}}&=-\frac{\Sigma \bar{\gamma}_1}{\Delta },\\
h_{t\phi}^{\mathrm{MGBK}} &= -\frac{\sin ^2\theta  \left[ a^2 M \gamma_{3,1}+a
   M^2 (\gamma_{1,2}+\gamma_{4,2})+M^3\gamma_{3,3}\right]}{\Sigma },\\
h_{\phi \phi}^{\mathrm{MGBK}} &=-\frac{\left(a^2+r^2\right)^2}{a^2} h_{tt} + \frac{\Delta \Sigma \bar{\gamma}_1}{a^2}-\frac{2 \left(a^2+r^2\right)}{a} h_{t\phi} - \frac{2 \Delta ^2 \bar{\gamma}_3 \sin ^2\theta}{a}+ \frac{2 \Delta ^2 \bar{\gamma}_4}{a^2},\\
\end{split}
\end{equation}
where
\begin{equation}
\begin{split}
P_1 &=-a^4 \left(r^2-a^2\right) \cos ^4\theta-2 a^2 \left(a^4+a^2 r^2+2 r^4\right) \cos^2\theta  +r^2 \left(2 a^4+5 a^2 r^2+r^4\right),\\
P_2 &= -a^4 \cos ^4\theta  (r-M) -a^2 \cos ^2\theta  \left[r^2 (2 M+r)-a^2 (r-2 M)\right] +r^2\left(2 a^2 M+a^2 r+M r^2\right),\\
P_3 &= a^4 \Delta  \cos ^4\theta  (r-2 M) +2 a^2 r^3 \cos ^2\theta  \left(a^2+4 M^2-2 M r+r^2\right)+r^3 (r-2 M) \left[a^2 (4 M+r)+r^2 (2 M+r)\right],\\
P_4 &= -2 a^2 r^2 \cos ^2\theta  \left(-a^2-2 M r+r^2\right) +r^3 \left[a^2 (3 r-4 M)+r^2 (r-2 M)\right]-a^4 \cos ^4\theta  \left(a^2-2 M r+3 r^2\right),\\
\end{split}
\end{equation}
\end{widetext}
and
\begin{equation}
\begin{split}
\hat{\rho}^2 &= r^2-a^2 \cos ^2\theta,\\
\hat{\Sigma} &= \left(a^2+r^2\right)^2-a^2 \Delta  \sin ^2\theta , \\
\bar{\gamma}_1&=\frac{\gamma_{1,2} M^2}{r^2} , \\
\bar{\gamma}_3&=\frac{\gamma_{3,1} M r^2+\gamma_{3,3} M^3}{r^4}, \\
\bar{\gamma}_4&=\frac{\gamma_{4,2} M^2}{r^2}.
\end{split}
\end{equation}
In the equations above, $\Sigma$ and $\Delta$ are the same as in the Kerr metric (see equation 2). Here we have resolved a typographical error present in the definition of $P_1$ in \citet{2011PhRvD..83j4027V}, which was also corrected in \citet{Vigeland2012} but not utilized in subsequent literature. The resolution of this error allows us to explore a wider range of values for the free parameters of this metric, $\gamma_{3,1},\,\gamma_{3,3},\, \gamma_{1,3},\, \mathrm{and} \gamma_{4,2}$, than was done in \citet{2013PhRvD..87l4017J}. We allow all free parameters to vary between $0.0$ and $2.0$.
In this form, the event horizon of this metric coincides with the Kerr horizon. This metric is also not Ricci flat.

\section{Simulating Shadows}\label{sec:methods}
We simulate a large set of black hole shadows that probe the allowed parameter space of the metrics described in the previous section. We perform ray-tracing simulations using the algorithm of \citet{Psaltis2012}, where we solve the geodesic equations in each metric to derive the black hole shadows. 

All of the metrics are axisymmetric, which allows us to use the two associated killing vectors to simplify the geodesic equations. We solve the simplified geodesic equations with a fourth-order Runge-Kutta integration scheme, with adaptive step size. We integrate the photon trajectories backwards starting at the observer's image plane and ending when the photon either comes within $10\%$ of the event horizon, or escapes to infinity. We denote by $x'$ and $y'$ the coordinates on the image plane (see Figure 1 in \citealt{2010ApJ...718..446J} for the geometry used for these simulations).  
For the QK metric we exclude the region inside $r=2.48\;M$ to avoid encountering any pathologies.

We describe the shape of a black-hole shadow in terms of its image-plane radius $R\equiv\sqrt{x'^2+y'^2}$ at different orientation angles $\psi = \cos^{-1}(x'/R)$.   We place 200 evenly spaced resolution elements  along $\psi$, for $0\le\psi\le2\pi$, and, for each value of this orientation angle, use the bisection method to find the boundary between photon trajectories that trace back into the black hole horizon and those that escape to infinity, i.e., the black-hole shadow; we terminate the bisection at an interval $\lesssim 10^{-6} M$. Since we use Boyer-Lindquist-like coordinates for all metrics, there is a pole that coincides with the spin axis of the black hole that often introduces numerical errors in the calculations. For this reason, we exclude orientations that are within $0.001$ radians of $\psi=\pi/2$ or $\psi=3\pi/2$ and interpolate between the nearby data points within these intervals.
\begin{figure}[h!]
\centering
\includegraphics[width=\columnwidth]{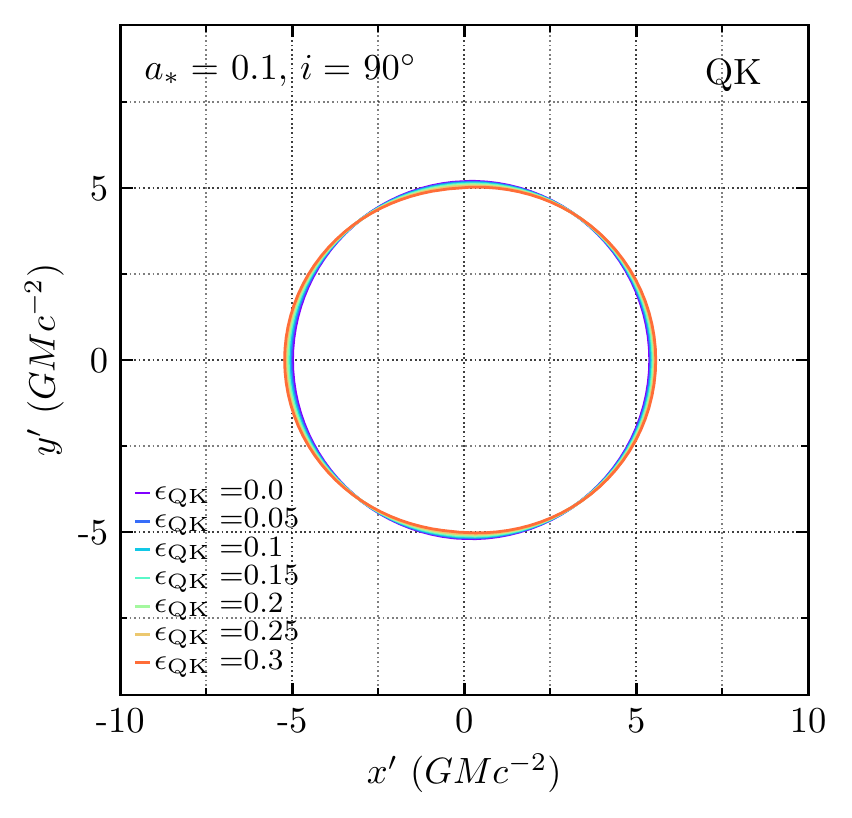}
\caption{Shadows that result from the QK metric as a function of the non-Kerr quadrupole $\epsilon_{\mathrm{QK}}$. Here, we have set the black-hole spin to $a_*=0.1$ and the observer's inclination to $i=90.0^{\circ}$. }
\label{fig:params_QK}
\end{figure}

In Figure~\ref{fig:params_QK}, we show how the shadow that results from the QK metric depends on the quadrupole deviation parameter $\epsilon_{\mathrm{QK}}$, while keeping the spin and the inclination angle of the observer constant. As discussed in~\citet{2010ApJ...718..446J}, the quadrupole deviation parameter in this metric introduces asymmetry to the shape of the shadow. However, because we opted here not to expand the geodesic equations in this metric to first order in $\epsilon_{\rm QK}$, we can only consider small values for this parameter before encountering metric pathologies. This results in shadow shapes that show only small deviations from the Kerr shadows.

\begin{figure*}[t!]
\centering
\includegraphics{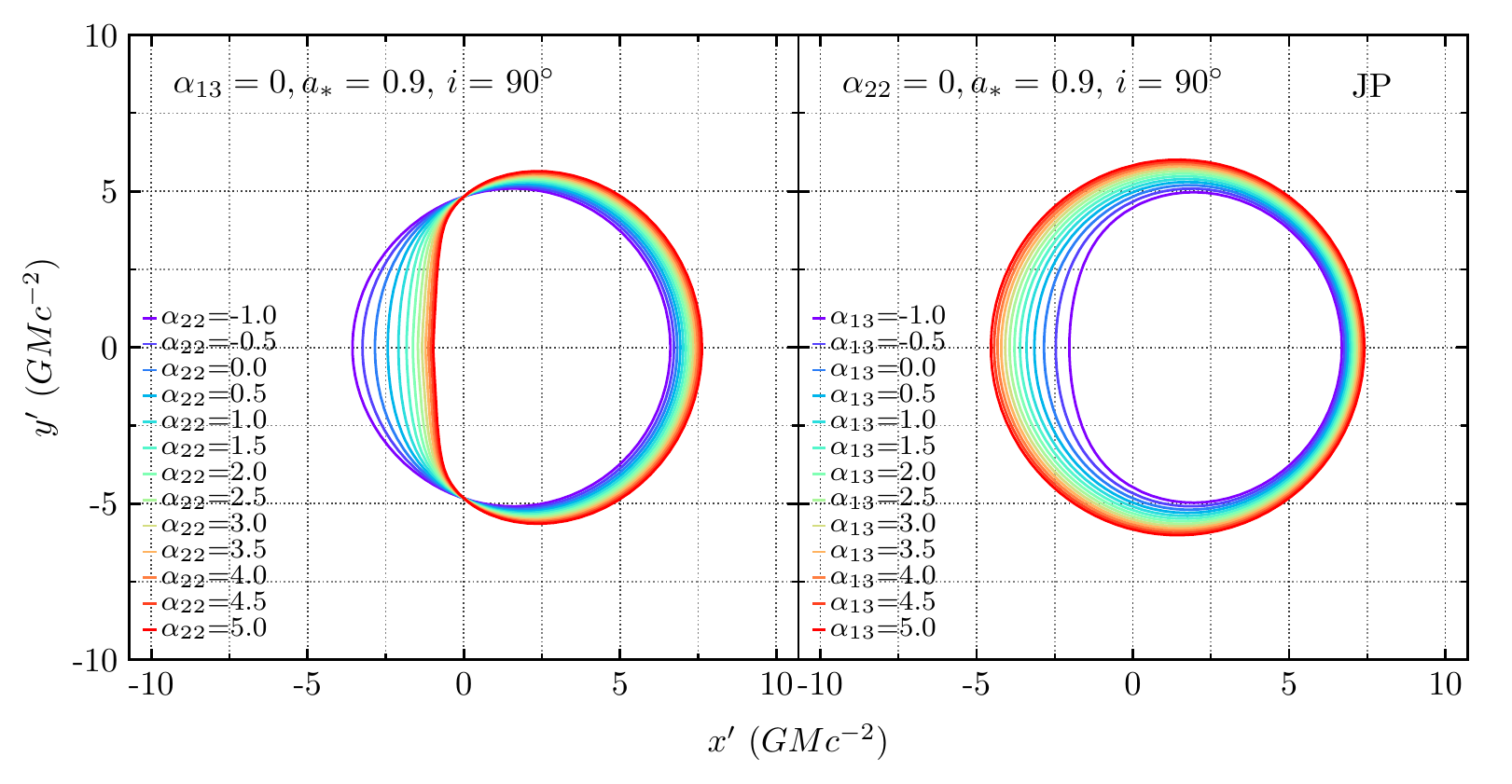}
\caption{Shadows that result from the JP metric as a function of the parameters $\alpha_{22}$ {\em (left)\/} and $\alpha_{13}$ {\em (right)\/}. Here, we have set the black-hole spin to $a_*=0.9$ and the observer's inclination to $i=90.0^{\circ}$. Additionally, we have set $\alpha_{13}=0$ for the left panel and $\alpha_{22}=0$ for the right panel.}
\label{fig:params_JP}
\end{figure*}

In Figure~\ref{fig:params_JP}, we show how the shadows that result from the JP metric depend on the deviation parameters $\alpha_{13}$ and $\alpha_{22}$. The effect of changing the parameter $\alpha_{22}$ on the shadow is similar to increasing the spin. For a Kerr metric, the shadow is approximately circular up to $a_*\approx 0.9$, after which a perturbation on the left side of the shadow appears (see Figure~\ref{fig:params_Kerr}). For the JP metric, increasing $\alpha_{22}$ increases the perturbation to the left side of the shadow far beyond the maximum for a Kerr black hole. 

On the other hand, the effect of changing the parameter $\alpha_{13}$ on the shadow is similar to increasing the black-hole mass. Kerr black hole shadows have radii between $4.8M$ and $5.2M$ for all spins. For the JP metric, increasing the parameter $\alpha_{13}$ tends to increase the size of the shadow, while making it more circular. Within the set of shadows we explored, the radius range for the JP metric is $5M\lesssim R\lesssim6 M$, indicating that a JP black hole with $\alpha_{13}=5$ can have a black hole shadow that is $\approx 20\%$ larger than a Kerr black hole of equivalent mass. 

\begin{figure*}[t!]
\centering
\includegraphics[width=.95\textwidth]{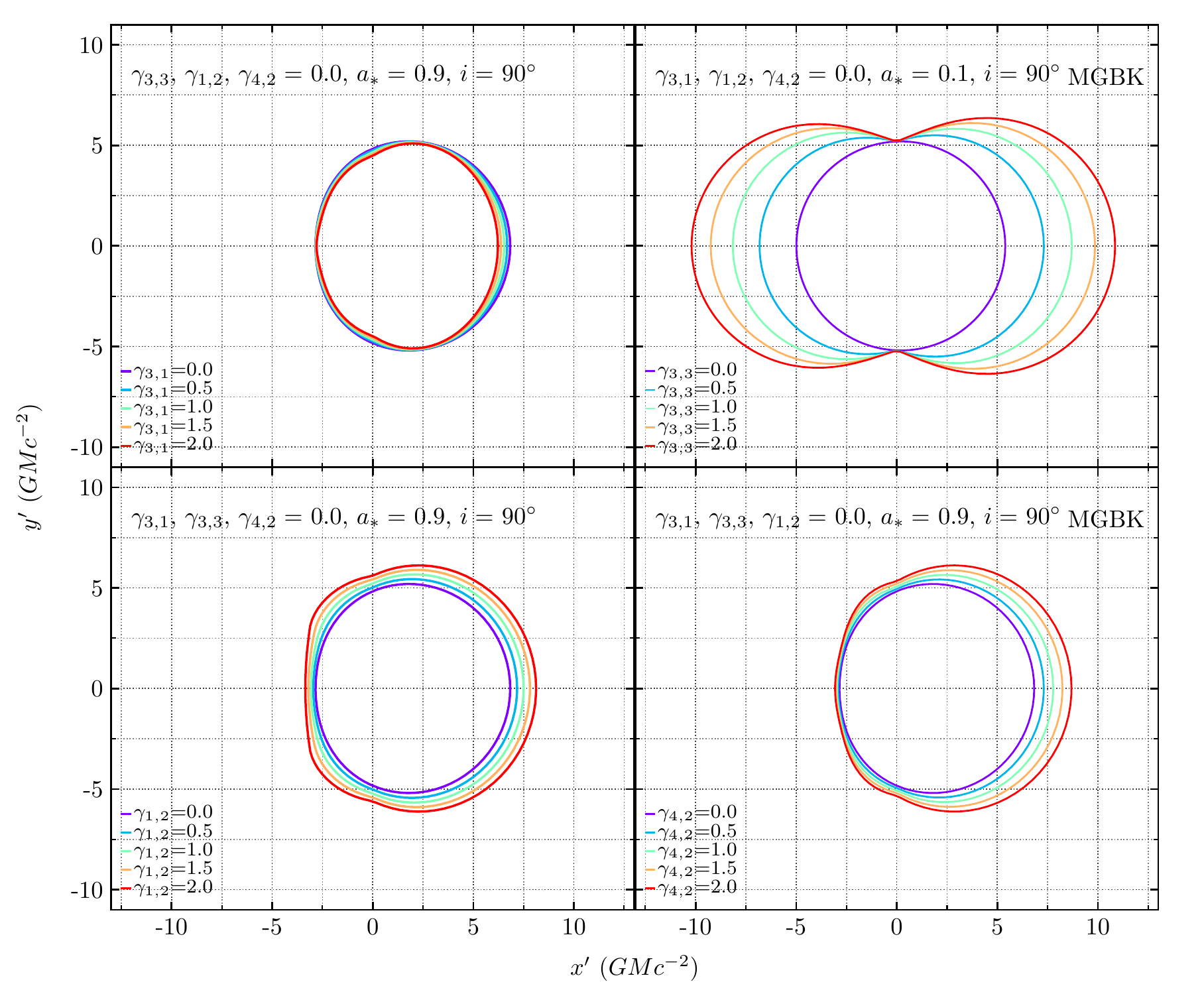}
\caption{Shadows that result from the MGBK metric as a function of the parameters $\gamma_{3,1}$ {\em (top left)\/},  $\gamma_{3,3}$ {\em (top right)\/}, $\gamma_{1,2}$ {\em (bottom left)\/} and $\gamma_{4,2}$ {\em (bottom right)\/}. Here, we have set the black-hole spin to $a_*=0.1$ in the top right panel and to $a_*=0.9$ in all other panels and the observer's inclination to $i=90.0^{\circ}$ in all panels. Additionally, we have set all perturbation parameters to zero except for the parameter that is varied in each panel.}
\label{fig:params_MGBK}
\end{figure*}

In Figure~\ref{fig:params_MGBK} we show how the shadows that result from the MGBK metric depend on the perturbation parameters $\gamma_{3,1},\, \gamma_{3,3},\,  \gamma_{1,2}$,\,and $\gamma_{4,2}$. The perturbations created by the four free parameters of the MGBK metric have some similarities. However, unlike the case of the other metrics, the shadows can look qualitatively different at low versus high $a_*$. For all parameters, the shadows seem to have ``dimples'' along the $x'=0$ line.

When $a_*$ is small, as is shown in the top right panel of Figure~\ref{fig:params_MGBK}, the shadows are relatively left-right symmetric and can become significantly elongated in the horizontal direction. The parameter $\gamma_{3,3}$ creates the most extreme shadows of this kind, as shown in the figure. In contrast, $\gamma_{3,1}$ has a negligible effect when $a_*$ is small. 

When $a_*$ is large, as is shown in the top left and bottom panels, the shadows become asymmetric with the right side becoming larger than the left due to the effect of frame dragging, as in the Kerr shadows in Figure \ref{fig:params_Kerr}.  The parameters $\gamma_{3,1}$ and $\gamma_{3,3}$ do not affect the overall size of the shadow for large $a_*$, while the parameters $\gamma_{1,2}$ and $\gamma_{4,2}$ can create larger shadows for large $a_*$ as shown. For high $a_*$ and high $\gamma_{1,2}$ shadows can become flat on the left side, similar to the behavior of Kerr shadows for very high spin.

\section{Centering and Rescaling Black Hole shadows}\label{sec:centering}

Frame-dragging effects cause a significant fraction of the shadows we have calculated to be displaced from the image origin that is centered on the black hole. Moreover, the effect of some perturbation parameters is simply to rescale the shadow sizes. Because our goal is to treat the set of shadows as an ensemble and compactly represent their shapes, we will first remove all trivial transformations. A simple translation or rescaling can be easily accounted for after the fact and does not need to be included as part of the PCA. By re-scaling and re-centering the shadows before analyzing their shapes, we can significantly reduce the number of parameters that will later be needed to fit to data. We emphasize that we do not wish to discard the information about the original size and displacement of the shadows but rather choose to consider it separately.

\begin{figure*}[p]
\centering
\includegraphics{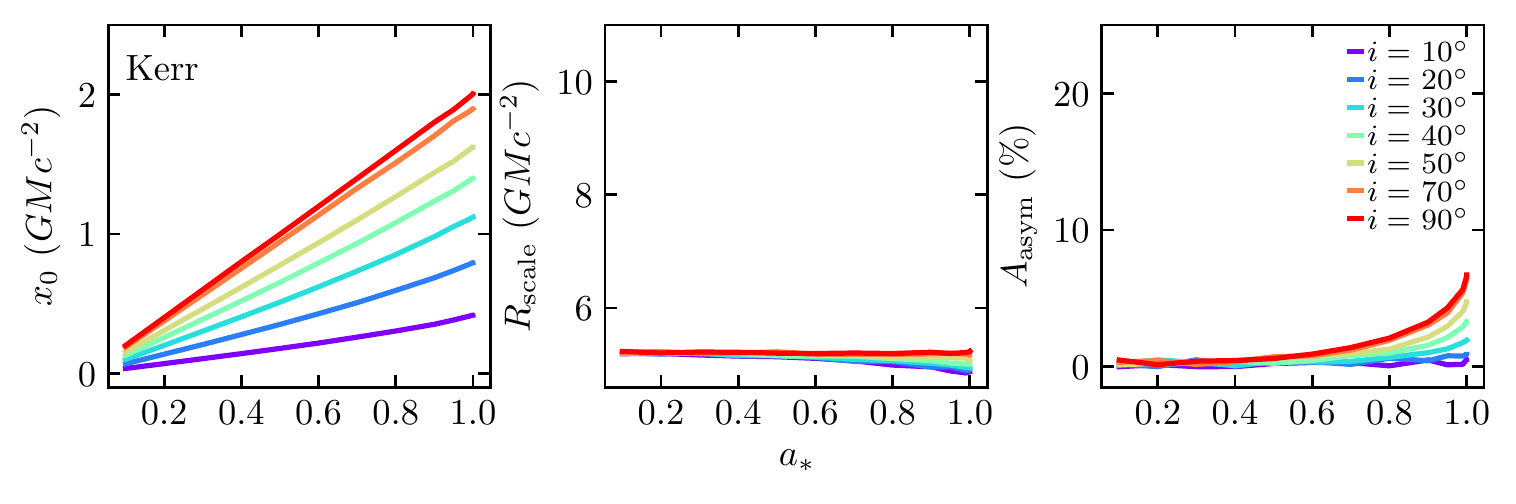}
\caption{The dependence of the displacement ($x_0$, left), the scaling factor ($R_{\mathrm{scale}}$, middle), and the asymmetry ($A_{\mathrm{asym}}$, right) on the spin of the black hole ($a_*$) for the Kerr metric. Different colors correspond to different values of the inclination angle, $i$. }
\label{fig:Kerr_x0_rscale}
\end{figure*}

\begin{figure*}[p]
\centering
\includegraphics{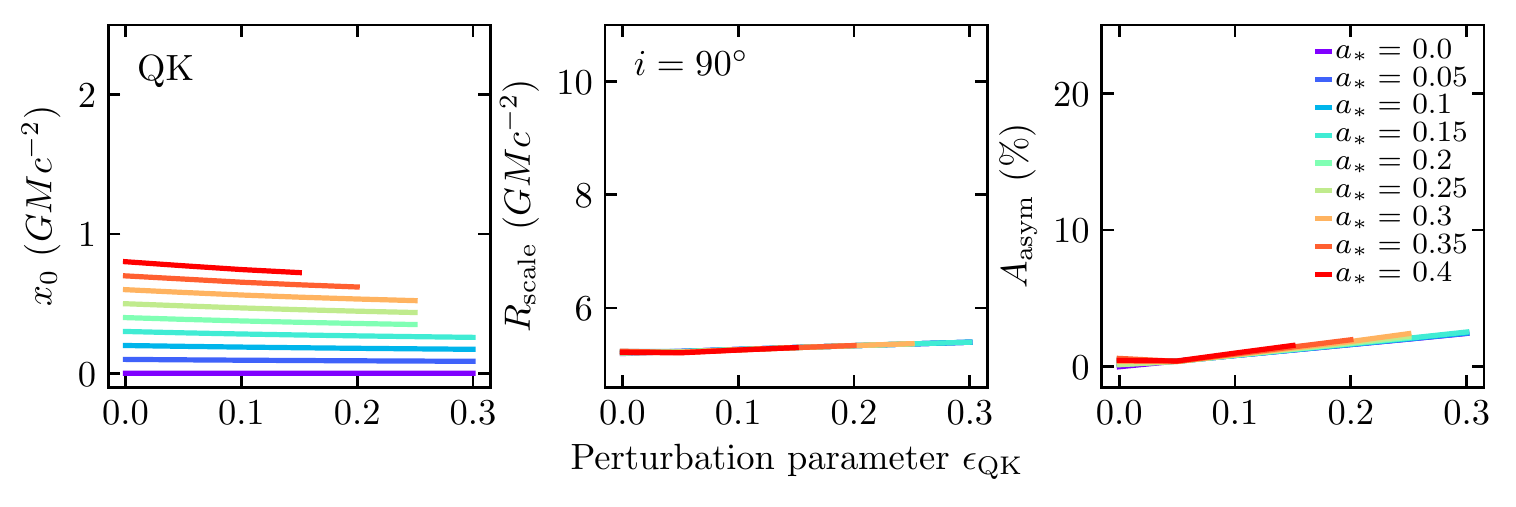}
\caption{The dependence of the displacement ($x_0$, left), the scaling factor ($R_{\mathrm{scale}}$, middle), and the asymmetry ($A_{\mathrm{asym}}$, right) on the perturbation parameter of the QK metric ($\epsilon_{\mathrm{QK}}$). Different colors correspond to different values of the black hole spin. }
\label{fig:QK_x0_rscale}
\end{figure*}

\begin{figure*}[p]
\centering
\includegraphics{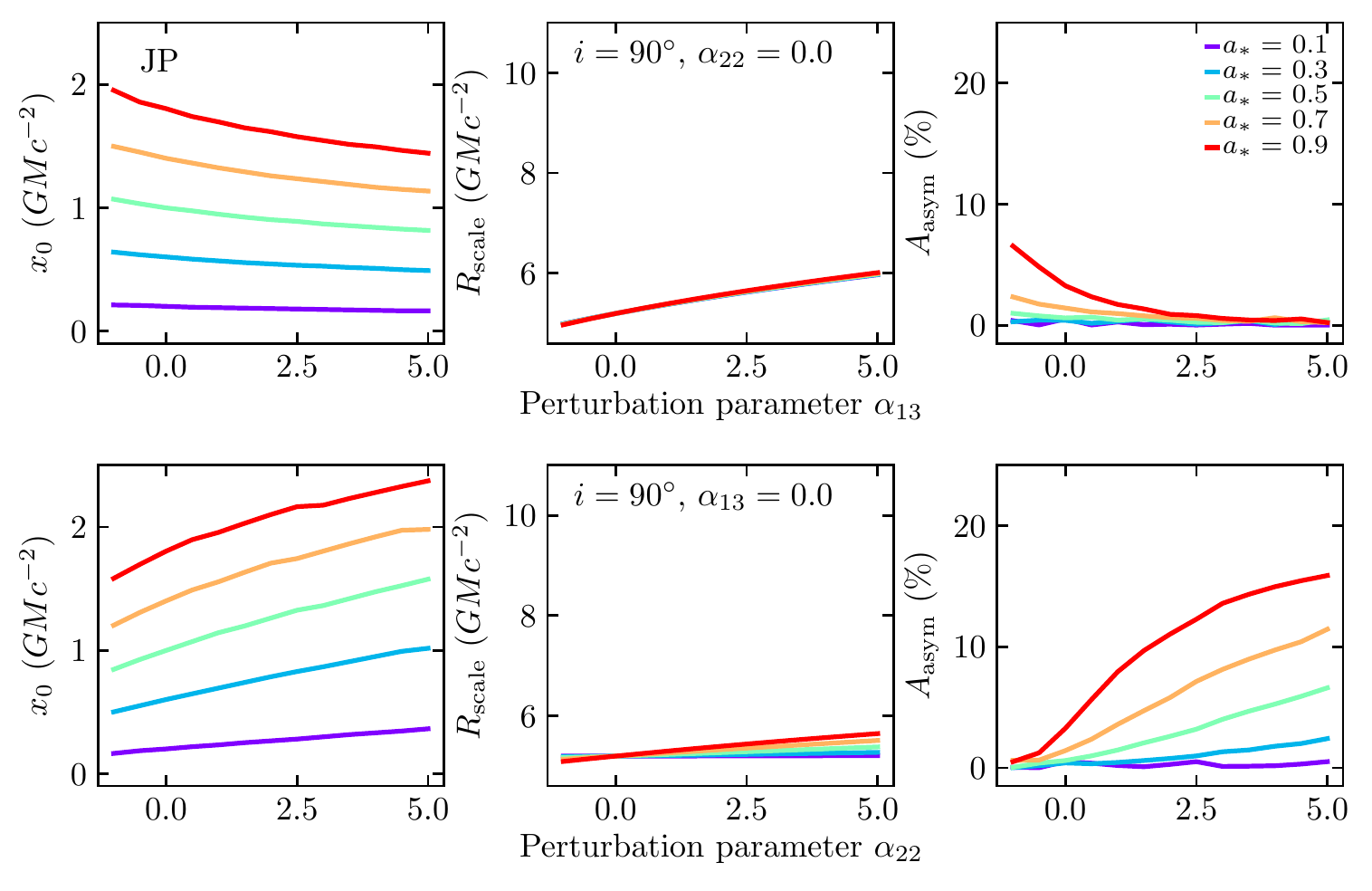}
\caption{The dependence of the displacement ($x_0$, left), the scaling factor ($R_{\mathrm{scale}}$, middle), and the asymmetry ($A_{\mathrm{asym}}$, right) on the perturbation parameters of the JP metric ($\alpha_{13}$ top row and $\alpha_{22}$ bottom row). Different colors correspond to different values of the black hole spin. 
}
\label{fig:JP_x0_rscale}
\end{figure*}

\begin{figure*}[t!]
\centering
\includegraphics{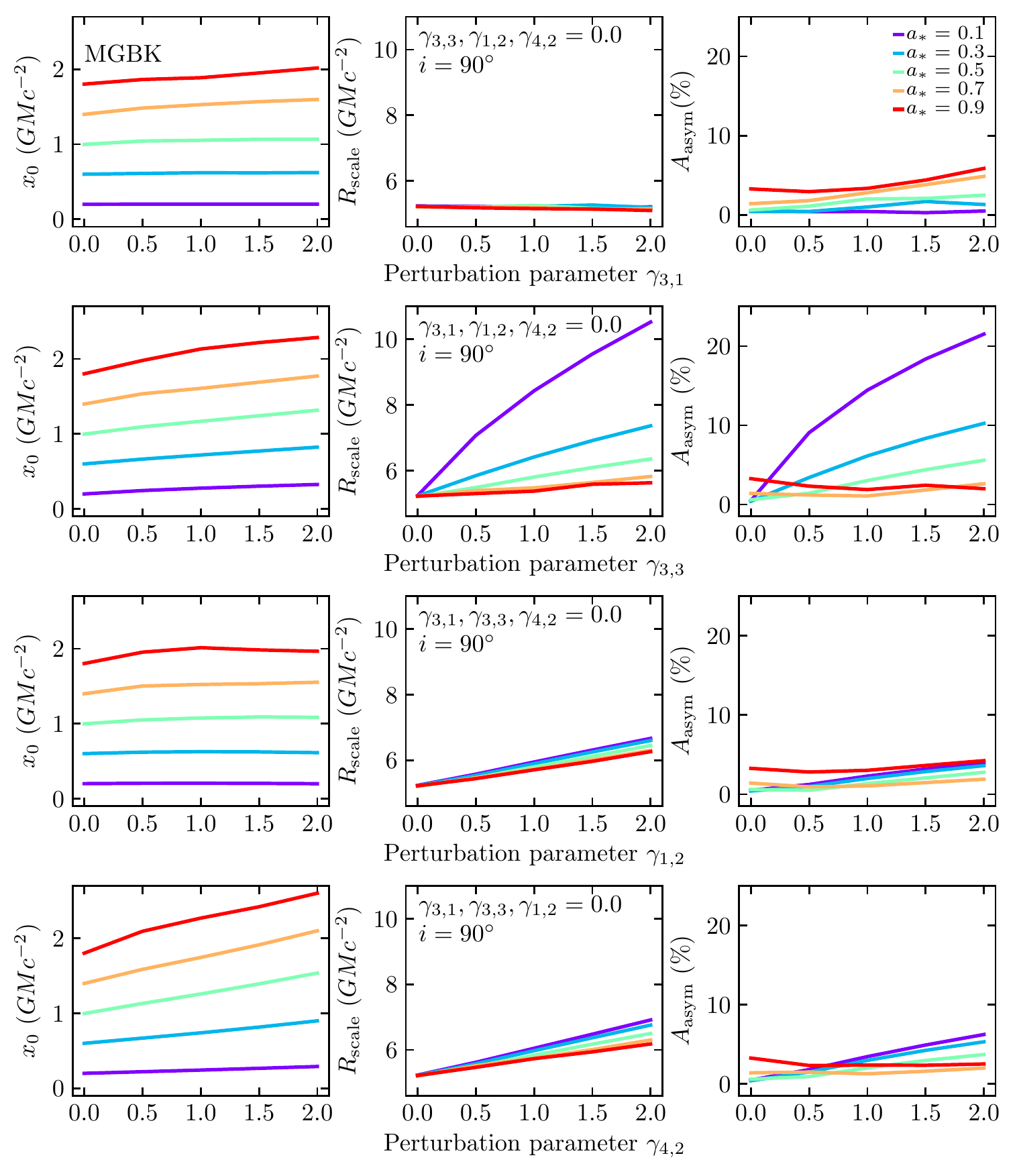}
\caption{The dependence of the displacement ($x_0$, left), the scaling factor ($R_{\mathrm{scale}}$, middle), and the asymmetry ($A_{\mathrm{asym}}$, right) on the perturbation parameters of the MGBK metric ($\gamma_{3,1}$ top row, $\gamma_{3,3}$ second row, $\gamma_{1,2}$ third row, and $\gamma_{4,2}$ bottom row). Different colors correspond to different values of the black hole spin.}
\label{fig:MGBK_x0_rscale}
\end{figure*}

There are many possible ways of defining the center of a perturbed circle, each resulting in a different set of parametrizations for the shadow shapes. For this work, we experimented with various methods and chose the method that resulted in the least number of PCA components (see \S \ref{sec:PCAintro}) necessary for accurate reconstructions. We also preferred a procedure for standardizing shadows that could be easily applied to the data as well as the simulations. We chose to re-center and re-scale each shadow such that it is as large as possible while still being contained within a circle centered at the origin with a radius equal to $\sqrt{27}M$, i.e., the radius of the shadow of a Schwarzschild black hole. 

We denote the amount by which each shadow is displaced in the $x'$ direction, i.e., perpendicular to the spin axis, by $x_0$, such that a shadow that is displaced to the right will have a positive $x_0$. Almost all of our shadows have negligible displacement in the $y'$ direction, i.e., along the spin axis, so we do not include a vertical displacement in this discussion. 

Before recording the shape of each shadow in terms of the image-plane radius $R$ along different orientation angles $\psi$ (see \S3), we first recenter it by subtracting $(x_0,\, y_0)$ from all coordinate pairs. We then define the scale of the shadow $R_{\rm scale}$ as the largest radius along the perimeter of the shadow and then rescale the shadows such that the largest radius becomes equal to $\sqrt{27}M$.

We further define the asymmetry ($A_{\mathrm{asym}}$) of a shadow as 
\begin{equation}
A_{\mathrm{asym}} = \frac{\sigma_{R}}{\bar{R}},
\end{equation}
where $\sigma_{R}$ is the standard deviation and $\bar{R}$ the mean of the radius of each shadow along different orientation angles $\psi$.

In Figures \ref{fig:Kerr_x0_rscale}, \ref{fig:QK_x0_rscale}, \ref{fig:JP_x0_rscale}, and \ref{fig:MGBK_x0_rscale} we show the dependence of the displacement ($x_0$), the scaling factor ($R_{\mathrm{scale}}$), and the asymmetry ($A_{\mathrm{asym}}$) of the shadows on the perturbation parameters of the various metrics and the black-hole spin;  for comparison, we also show the same parameters for the Kerr metric. As expected, for all metrics, the shadows are displaced to the right with increasing black hole spin, with the perturbation parameters having only a secondary effect. At the same time, changing most of the perturbation parameters introduces changes in the overall scale $R_{\rm scale}$ of the shadows to amounts larger than the $\pm 4$\% range obtained for the Kerr metric. Both the JP and the MGBK metrics can create asymmetries, $A_{\rm asym}$, that are much larger than what is seen for the Kerr metric. 
 
\begin{figure}[t!]
\centering
\includegraphics[width=\columnwidth]{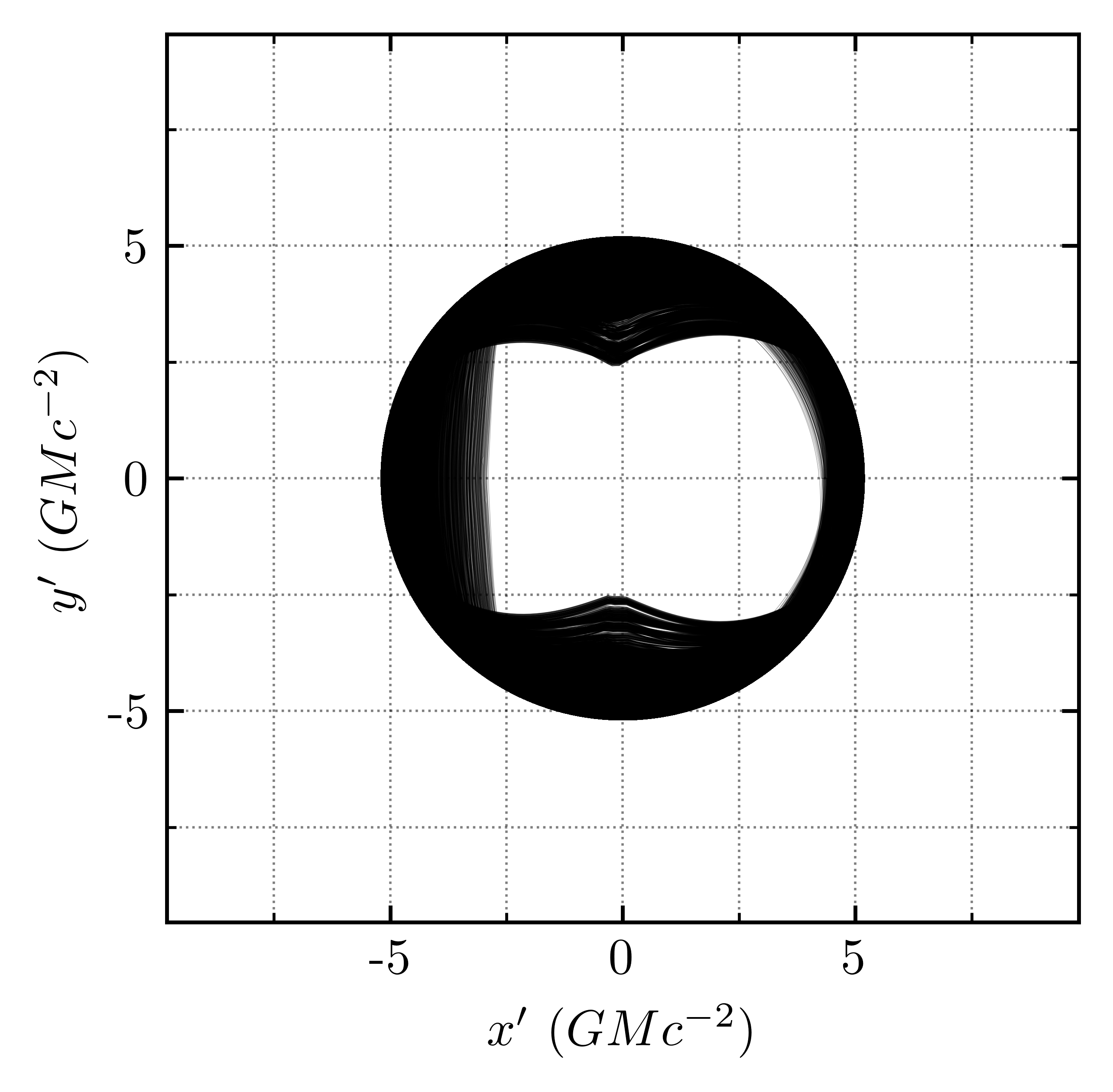}
\caption{The shaded region shown contains all shadows in the ensemble after they have been re-scaled and re-centered.}
\label{fig:circles_all}
\end{figure}

Before performing PCA on the re-centered and re-scaled set of shadows, we interpolate each shadow so that they are all evaluated on evenly spaced values of $\psi$ and aligned with all other shadows. In Figure~\ref{fig:circles_all} we show the full set of shadows after they have been re-scaled and re-centered. A total of 23,887 shadows are included in the set: 15,625 for the MGBK metric, 7,605 for the JP metric, 522 for the QK metric, and 135 for the Kerr metric. For the QK metric, we allowed the spin of the black hole to vary between $0.0$ and $0.4$ in intervals of $0.05$ and the perturbation parameter $\epsilon_{\mathrm{QK}}$ to vary between $0.0$ and 0.35 in intervals of 0.05. (As previously noted, the limit of allowed values of the perturbation parameter $\epsilon_{\mathrm{QK}}$ is smaller for certain combinations of $a_*$ and $i$.) For the JP metric, we allowed the spin of the black hole to vary between $0.1$ and $0.9$ in intervals of $0.2$, and both of the perturbation parameters, $\alpha_{13}$ and $\alpha_{22}$, to vary between $-1$ and $5$ in intervals of 0.5. For the MGBK metric we allowed $a_*$ to vary from $0.1$ to $0.9$ in intervals of $0.2$ and allowed the perturbation parameters, $\gamma_{3,1},\,\gamma_{3,3},\,\gamma_{1,2},\,$ and $\gamma_{4,2}$, to vary between $0.0$ and $2.0$ in intervals of $0.5$. For the Kerr metric we allowed $a_*$ to vary between $0.1$ and $0.9$ in intervals of $0.1$ and also included the values $0.95, \,0.99,\, 0.9999,\, 0.99999,\, 0.999999,\, \mathrm{and}\,0.99999999$. For all metrics except MGBK, we allowed the inclination angle $i$ to vary between $10^{\circ}$ and $90^{\circ}$ in intervals of  $10^{\circ}$; for MGBK we used intervals of $20^{\circ}$.

\section{Principal Components Analysis}\label{sec:PCAintro}

In order to compare the shadow shapes we have simulated to observations, we would like to express their shapes using a small number of parameters that can be directly inferred from the data. \citet{2015MNRAS.454.2423A} followed a similar approach by representing the shadows as a sum of Legendre polynomials. Here we explore the use of Principal Components Analysis (PCA) to compactly represent the space spanned by Kerr and non-Kerr shadows. 

The PCA algorithm diagonalizes the covariance matrix of the data to find a set of orthogonal basis vectors (principal components) that are ordered such that the first principal component accounts for the largest possible percentage of the variance in the data set, the second principal component accounts for the second largest percentage of the variance, and so on (see e.g., \citealt{2010AJ....140..390B}). For data sets that are correlated, this can result in significant dimensionality reduction since it may be possible to reconstruct the original data set using only a handful of PCA components.

We represent each shadow as a column vector $\textbf{S}_n\equiv S_{in}$, which corresponds to the radius of the $i$-th orientation point of the $n$-th shadow. In addition to re-scaling and re-centering as was discussed above, we also subtract from each shadow shape the constant $\sqrt{27}M$, such that all circular shadows would have a radius of zero after re-scaling, re-centering, and subtraction.

We define the covariance matrix $C$ as
\begin{equation}
C = \frac{1}{m}\sum_{n=1}^{m} (\textbf{S}_n-\sqrt{27}M)(\textbf{S}_n-\sqrt{27}M)^T,
\end{equation}
and find a basis of eigenvectors, or ``eigenshadows", by diagonalizing $C$ such that
\begin{equation}
C \textbf{u}_k = \lambda_k \textbf{u}_k, 
\end{equation}
where $\textbf{u}_k$ is the $k$-th eigenshadow, $\lambda_k$ is the $k$-th eigenvalue, and $k=1,2, ... ,m$. Although PCA returns $m$ eigenshadows, they are ordered such that only the first few eigenshadows are needed to reconstruct the majority of the variance within the data set. The associated eigenvalue of an eigenshadow indicates the percentage of the variance within the data set that this particular eigenshadow accounts for.
Because of this, we normalize each eigenshadow such that
\begin{equation}
\textbf{u}_k^2 = \lambda_k,
\end{equation}
and normalize the eigenvalues such that
\begin{equation}
\sum^m_{k=1}\lambda_k=1.
\end{equation}

Once we have derived a basis of eigenshadows, we can reconstruct each shadow as a linear combination of the eigenshadows. Specifically, we write
\begin{equation}
\textbf{S}_n = \sqrt{27}M + \sum^m_{k=1} a_{nk} \textbf{u}_k, 
\end{equation}
where the amplitude $a_{nk}$ is the coefficient multiplying the $k$-th eigenshadow in the linear combination of the $n$-th shadow. The overall sign of each eigenshadow is arbitrary and any eigenshadow can have both positive and negative coefficients ($a_{nk}$). For simplicity, we choose the sign of each eigenshadow such that the amplitude with the largest magnitude is positive. In the next section we will also discuss the relative contribution of eigenshadow $k$ to a shadow $n$, which we define as 
\begin{equation} 
a'_{nk} =\frac{a_{nk} \sigma_{\textbf{u}_k} }{\sqrt{27}M},
\end{equation}
where  $\sigma_{\textbf{u}_k}$ is the standard deviation of the $k$-th eigenshadow defined as
\begin{equation} 
\sigma_{\textbf{u}_k} = \sqrt{\frac{1}{N} \sum^{N}_{i=1} \left( u_{ik} - \langle \textbf{u}_{k}\rangle \right)^2}\;.
\end{equation}
Here, $u_{ik}$ is the $i$-th element of the $\textbf{u}_k$ eigenshadow, and $\langle \textbf{u}_{k}\rangle$ is the average of the elements of the $\textbf{u}_k$ eigenshadow. We will also refer to $a'_{nk}$ as the rescaled amplitude.

In principle, to reconstruct a shadow to high accuracy, we would need all $m$ eigenshadows. However, because PCA components are ordered by their relative importance, the first few eigenshadows can already reconstruct the large scale features of each shadow, as long as the eigenvalues decline reasonably fast. This allows for significant dimensionality reduction as only a handful of eigenshadows may be needed depending on the desired accuracy. In the following sections, we will explore how many eigenshadows are needed for accurate reconstructions.

\begin{figure}[t!]
\centering
\includegraphics[width=\columnwidth]{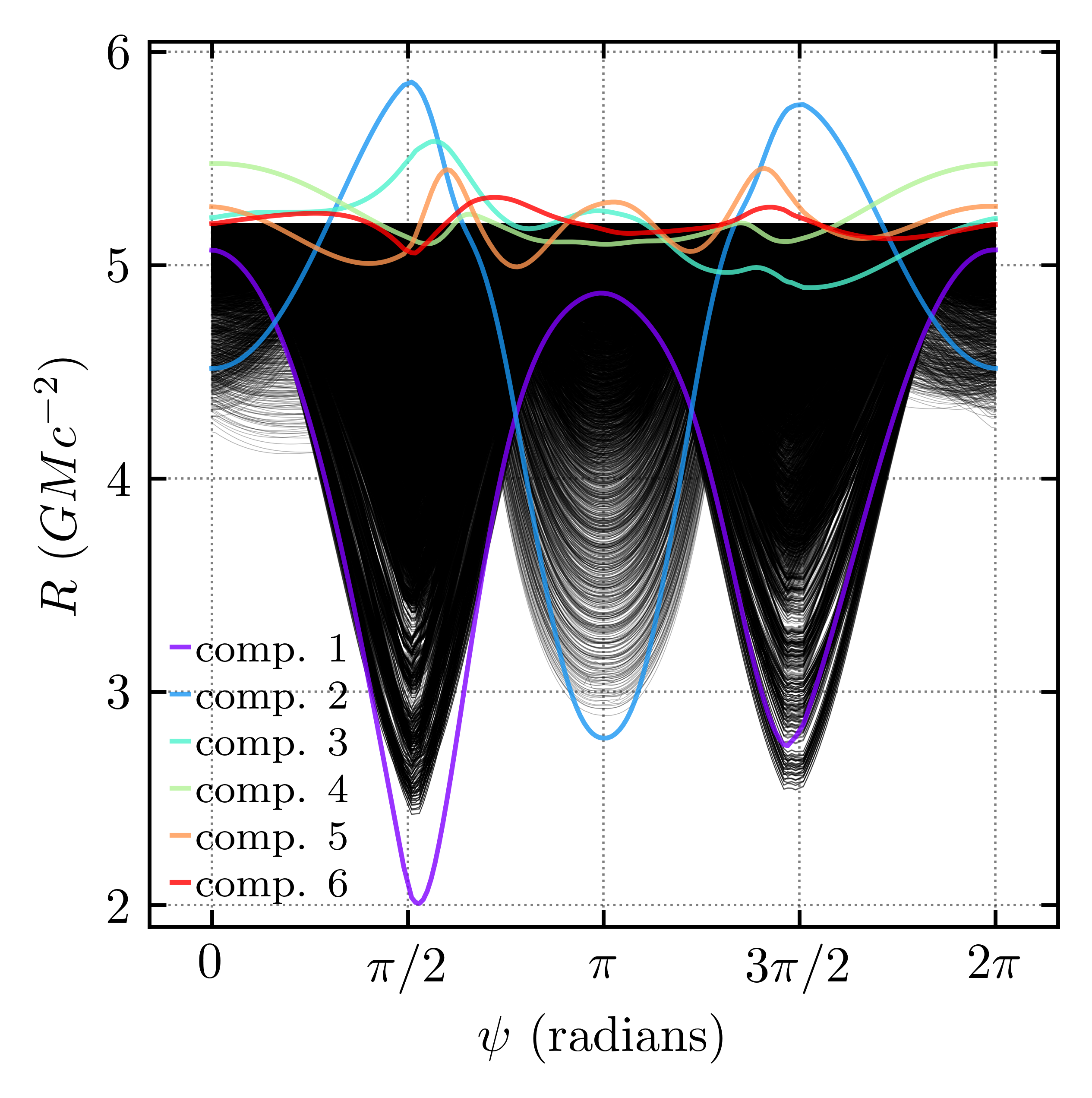}
\caption{All re-centered and re-scaled shadows in the ensemble after they have been unwrapped into curves of radius $R$ as a function of the orientation angle $\psi$. The first six PCA components  are also shown as colored curves, each multiplied by the largest magnitude amplitude ($a_{nk}$) for each component. }
\label{fig:lines_all}
\end{figure}

\begin{figure}[t!]
\centering
\includegraphics[width=\columnwidth]{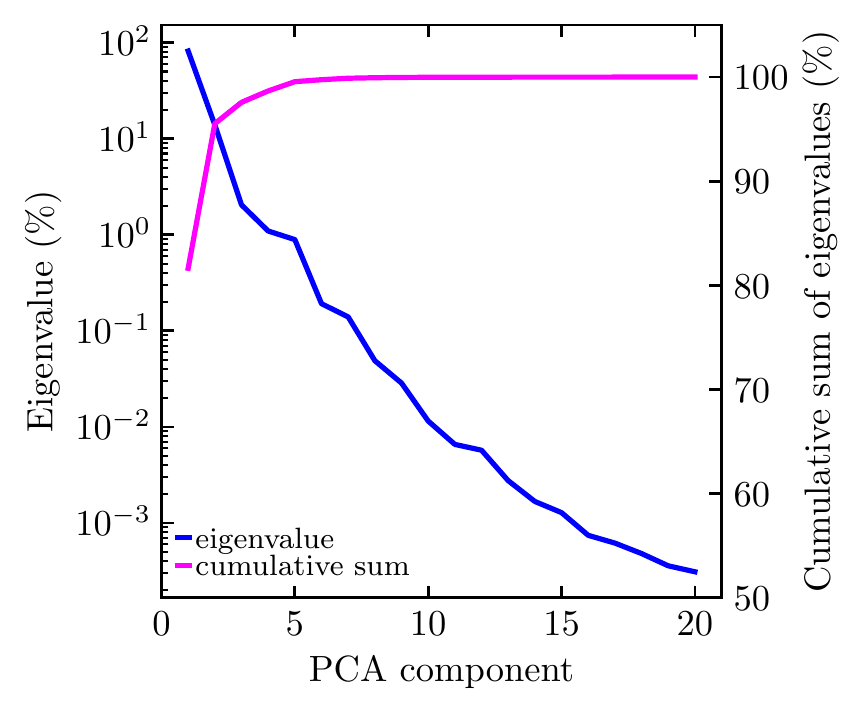}
\caption{The spectrum of the PCA eigenvalues {\em (blue)\/} and the cumulative sum of the eigenvalues  {\em (magenta)\/}. The PCA eigenvalues quantify the percentage of the overall variance in the data set that each PCA component accounts for. The rapid decline in eigenvalue for the higher components indicates that PCA can be used for efficient dimensionality reduction. The first component accounts for 81.6\% of the variance and only five PCA components can account for 99.6\% of the variance. }
\label{fig:spectra}
\end{figure}

\begin{figure*}[t!]
\centering
\includegraphics[width=\textwidth]{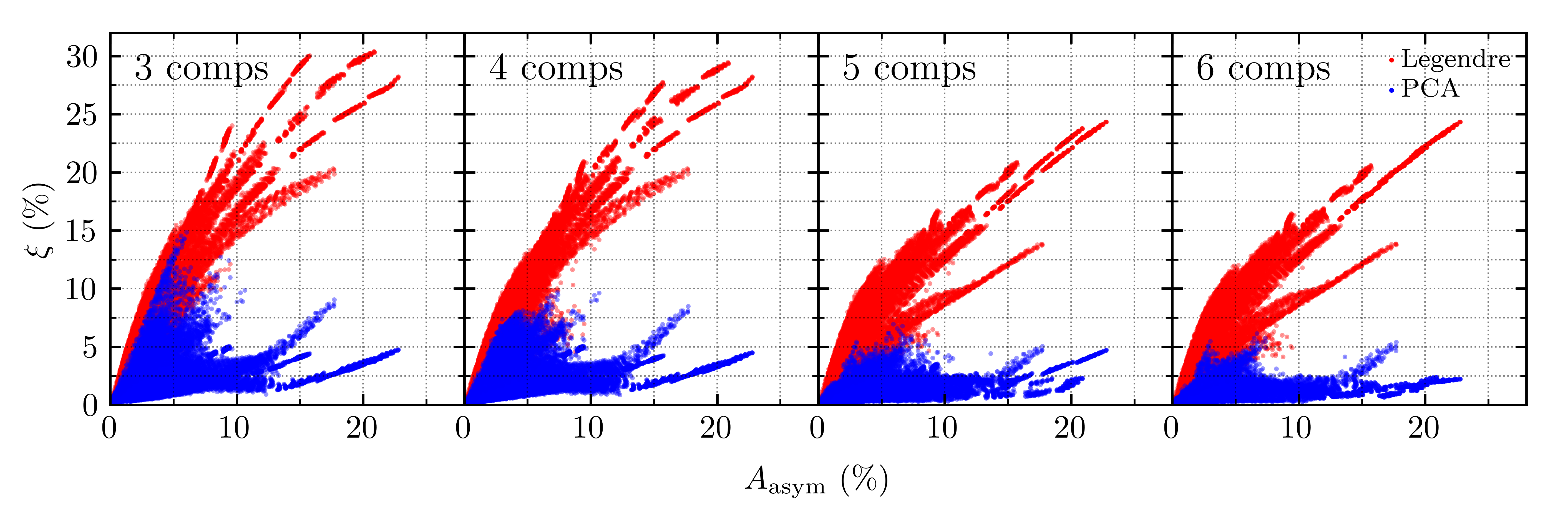}
\caption{{\em (Blue points)\/} The maximum fractional difference between the shape of a shadow in our ensemble and its reconstruction using three to six (panels left to right) PCA components plotted against the degree of asymmetry of each shadow. {\em (Red points)\/} The maximum fractional difference between the shape of a shadow and its reconstruction using three to six Legendre polynomials. The PCA components provide a representation of shadow shapes that is more compact and efficient than using Legendre polynomials, with a maximum error of $\lesssim 5$\% for reconstructions with five PCA components.}
\label{fig:errors_compare}
\end{figure*}

In Figure~\ref{fig:lines_all}, we show the full set of shadows unwrapped into curves of radius $R$ as a function of the orientation angle $\psi$. We also show the first six PCA components that were derived from the set of shadows. In this figure, we have multiplied each PCA component by the largest  amplitude needed to reconstruct the shadows within the set, effectively showing the largest possible contribution that each eigenshadow has to any shadow within the set. 

In Figure~\ref{fig:spectra}, we show the spectrum of PCA eigenvalues. The rapid decrease in PCA eigenvalues indicates that only a small number of PCA components are needed to reproduce the majority of the variance in the data set. Specifically, the first PCA component accounts for 81.6\% of the variance in the data set, the second component accounts for 13.9\%, and only four PCA components are needed to account for 98.6\% of the variance.

\section{PCA reconstructions}\label{sec:PCArecon}

In this section we quantify the accuracy of the PCA reconstructions using only a small number of components and contrast it to reconstructions using Legendre polynomials (as done in, e.g., \citealt{2015MNRAS.454.2423A}). Additionally, we explore the relation between the amplitudes of the PCA components needed to reconstruct a particular shadow and the parameters of the underlying metrics. 

We define the maximum fractional difference ($\xi$) for the PCA reconstruction of each shadow as 
\begin{equation}
 \xi= \frac{\mathrm{max} (|S_{in} - S_{in}'|)}{\sqrt{27}M}, 
\end{equation}
where $S_{in}$ denotes the complete shadow and $S_{in}'$ denotes the reconstructed shadow. We apply this metric to reconstructions with either PCA components or with Legendre polynomials. For consistency with the PCA reconstructions we perform the Legendre polynomial reconstructions on the re-centered and re-scaled shadows. 

In Figure~\ref{fig:errors_compare} we plot the maximum fractional difference as a function of the asymmetry of each shadow for reconstructions using 3 to 6 PCA components and 3 to 6 Legendre polynomials. As expected, both methods have similar performance for shadows with small asymmetry. However, the maximum fractional difference in the Legendre polynomial reconstructions is significantly higher than that of the PCA reconstruction for shadows with higher asymmetry. A reconstruction with only five Legendre polynomials for shadows with a $\sim 20$\% asymmetry will lead to a fractional difference as high as 25\%. In contrast, only five PCA components are needed to reconstruct all shadows within the ensemble with $\lesssim 5$\%  maximum fractional error. Adding a sixth PCA component does not substantially improve the reconstruction.

The shadows that have the highest ($\gtrsim 5\%$) fractional maximum error in reconstructions are MGBK shadows with high perturbation parameters and high values of spin, such as those shown in the bottom panels of Figure~\ref{fig:params_MGBK}. Most of these shadows have small ``dimples'' at $x'=0$ before they are aligned, which are displaced to different $xÕ$ locations, depending on spin, after alignment. These dimples make the shadows particularly hard to reconstruct because their varying locations cannot be captured by a small number of PCA components.

\begin{figure*}[t!]
\centering
\includegraphics{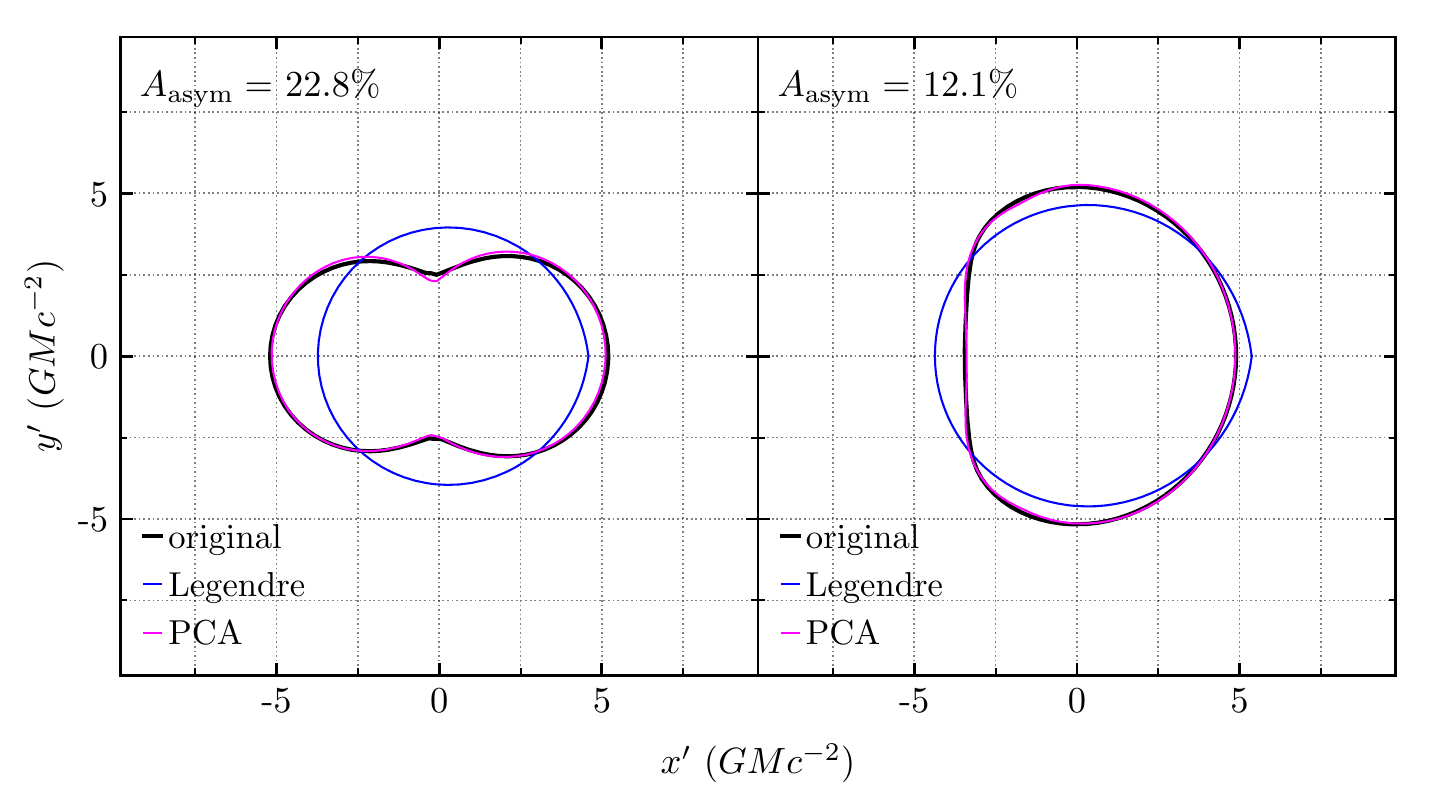}
\caption{Reconstructions of two highly asymmetric black-hole shadows (black curves) with 4 PCA components (magenta curves) and 4 Legendre polynomials (blue curves). The left panel shows a shadow created with the MGBK metric with an asymmetry of $23\%$. The right panel shows a shadow created with the JP metric with an asymmetry of $12\%$. Using a small number of PCA components generates  reconstructions of black hole shadows that are of higher fidelity compared to using an equal number of Legendre polynomials.}
\label{fig:recon_hard}
\end{figure*}

In Figure~\ref{fig:recon_hard} we show two sample reconstructions for shadows with high asymmetry using either four PCA components or four Legendre polynomials. The left panel corresponds to the MGBK metric with parameters $a_*=0.1$, $i=90^{\circ}$,  $\gamma_{3,1}=0.0$, $\gamma_{3,3}=2.0$, $\gamma_{1,2}=0.0$, $\gamma_{4,2}=2.0$ and the right panel corresponds to the JP metric with parameters $a_*=0.9$, $i=70^{\circ}$,  $\alpha_{13}=2.5$, $\alpha_{22}=5.0$. Only a small number of PCA components are needed to generate high fidelity reconstructions. Reconstructions with the same number of Legendre polynomials result in significantly higher errors.

\begin{figure*}[t!]
\centering
\includegraphics{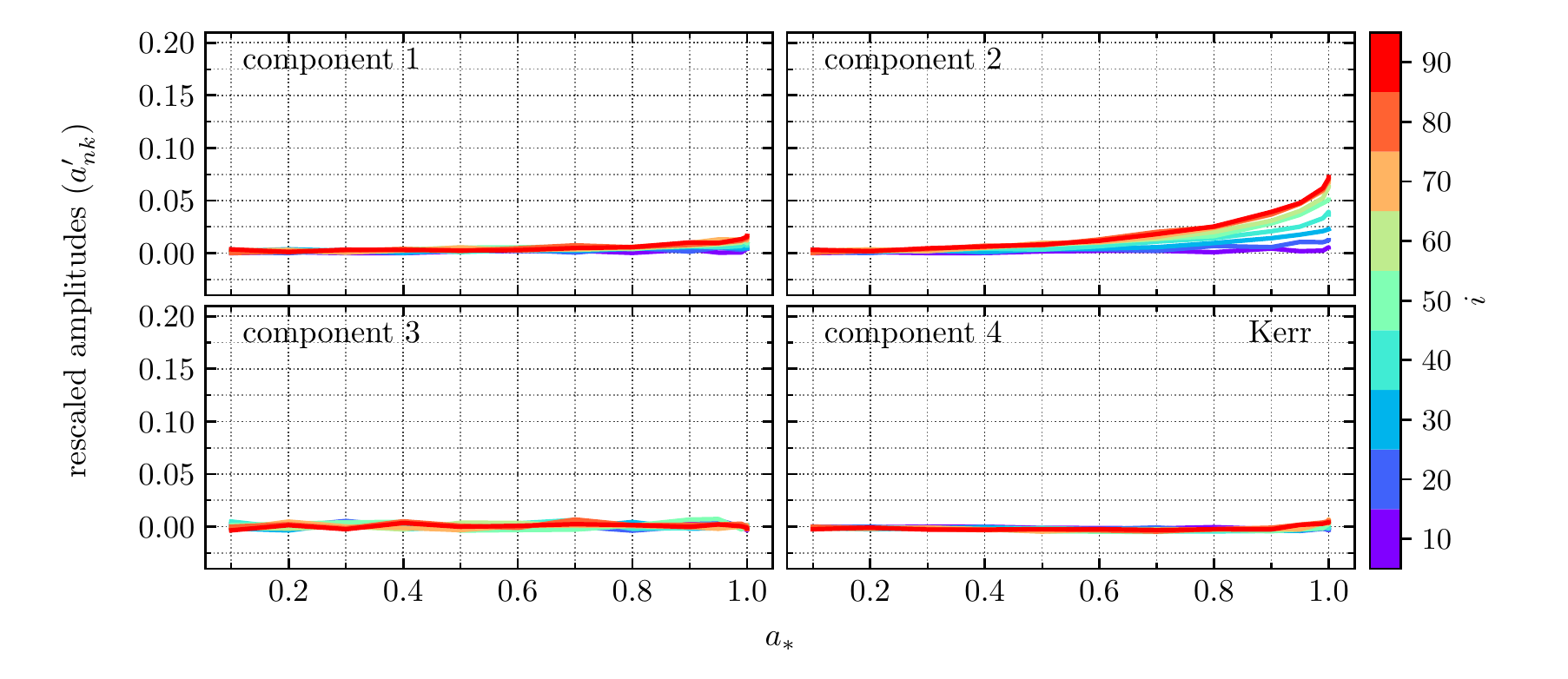}
\caption{The relation between the rescaled amplitude ($a'_{nk}$) of the PCA components of shadows calculated with the Kerr metric and the black hole spin ($a_{\mathrm{bh}}$). Different colors correspond to different observer inclinations ($i$).}
\label{fig:amps_vs_gamKerr}
\end{figure*}
\section{Discussion}\label{sec:conclusion}

We have calculated an ensemble of over 20,000 simulated black hole shadows that probe the allowed space of several parametrized non-Kerr metrics. We applied PCA to our ensemble of shadows and derived an orthogonal basis for shadow shapes.  We then used this analysis to reduce the dimensionality of the set of eigenshadows and showed that only five PCA components are required to reproduce all shadows within the ensemble, with a maximum fractional difference ($\xi$) of $\lesssim 5\%$. We compared the accuracy of the PCA reconstructions to reconstructions using Legendre polynomials and found that, for shadows with high asymmetry, the PCA reconstructions introduce errors that are nearly ten times lower than the Legendre polynomial reconstructions. 

\begin{figure*}[t!]
\centering
\includegraphics{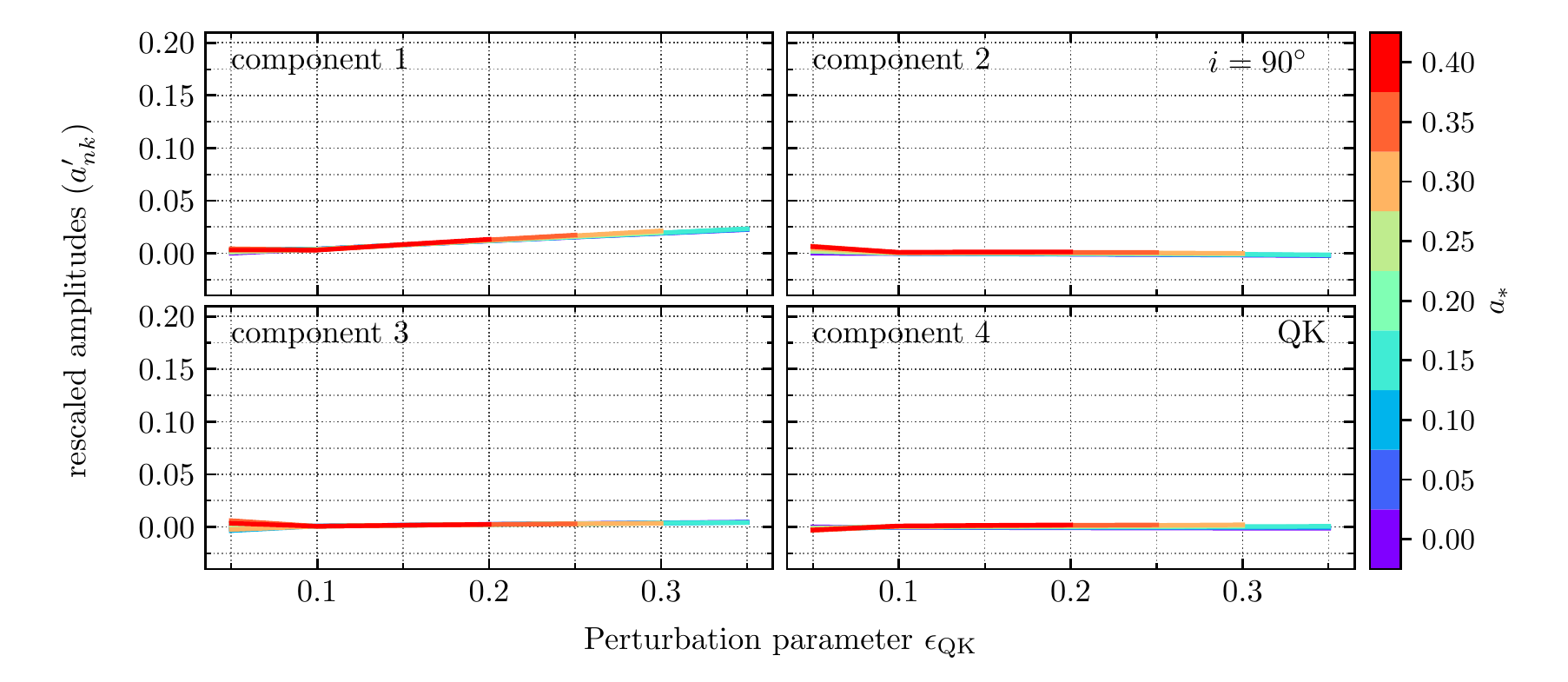}
\caption{The relation between the rescaled amplitude ($a'_{nk}$) of the PCA components of shadows calculated with the QK metric and the perturbation parameter $\epsilon_{\rm QK}$. Here we have set the observer's inclination to $i=90.0^{\circ}$ and different colors correspond to different black hole spins.}
\label{fig:amps_vs_gamQK}
\end{figure*}

\begin{figure*}[t!]
\centering
\includegraphics{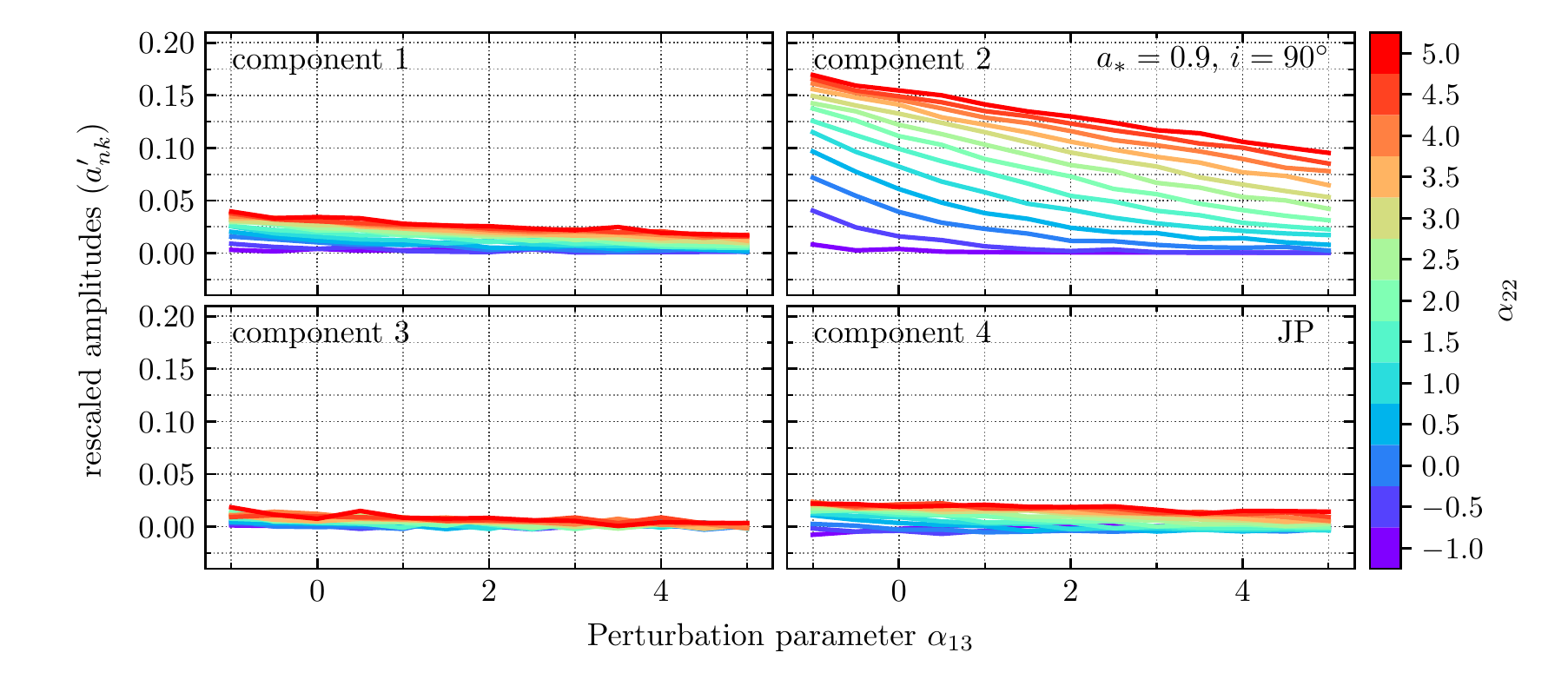}
\caption{ Same as Figure~\ref{fig:amps_vs_gamQK} but for the shadows and perturbation parameters of the JP metric.  The different colors correspond to different values of the perturbation parameter $\alpha_{22}$.}
\label{fig:amps_vs_gamJP1}
\end{figure*}

\begin{figure*}[t!]
\centering
\includegraphics{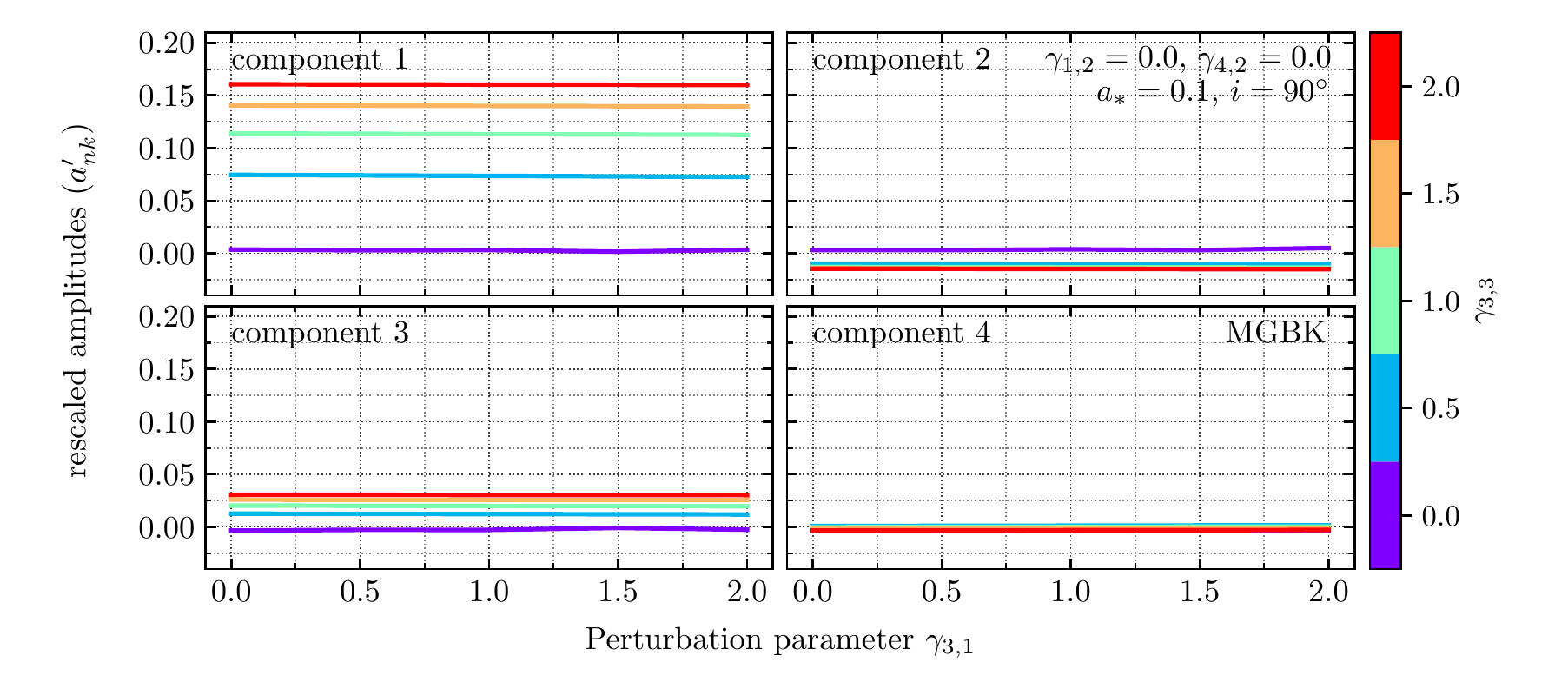}\\[-2ex]
\includegraphics{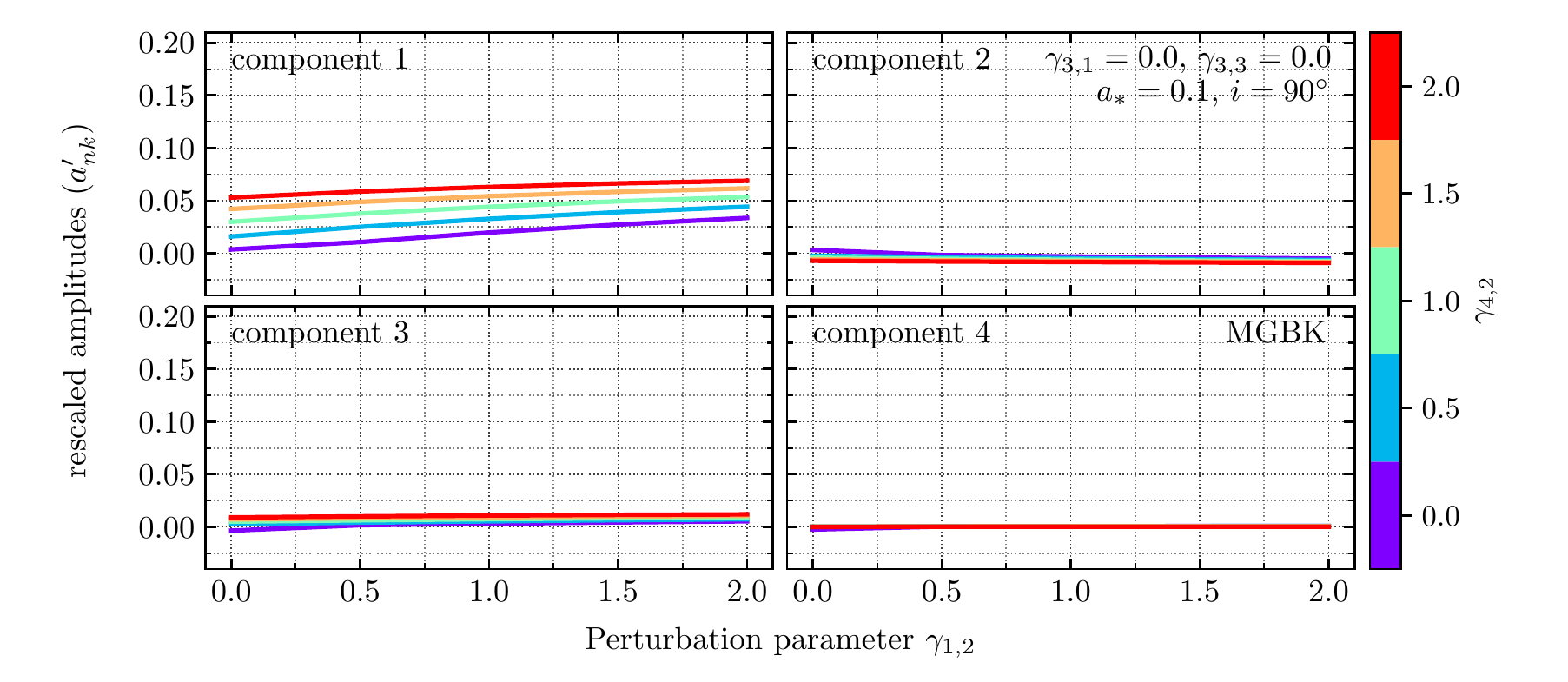}
\caption{ Same as Figures \ref{fig:amps_vs_gamQK} and \ref{fig:amps_vs_gamJP1} but for the shadows and perturbation parameters of the MGBK metric. Here we have set the black-hole spin to $a_{\mathrm{bh}}=0.1$ and the observer's inclination to $i=90.0^{\circ}$. In the top four panels, the different colors correspond to different values of the perturbation parameter $\gamma_{3,3}$ and, in the bottom four panels, the colors correspond to different values of $\gamma_{4,2}$. }
\label{fig:amps_vs_gamMGBK1}
\end{figure*}

This approach allows us to measure or place constraints on deviations from the Kerr metric using black hole images. Given a set of measurements and choosing a particular metric, we can convert the measured coefficients of the PCA components to metric parameters. Figures \ref{fig:amps_vs_gamKerr}, \ref{fig:amps_vs_gamQK}, \ref{fig:amps_vs_gamJP1},  and \ref{fig:amps_vs_gamMGBK1}  show sample mappings between the rescaled amplitudes ($a'_{nk} $) of the PCA reconstructions of shadows and the parameters of the underlying metrics.  Such mapping will allow us to place metric-specific constraints on deviations from the Kerr metric.

\begin{figure}[t!]
\centering
\includegraphics{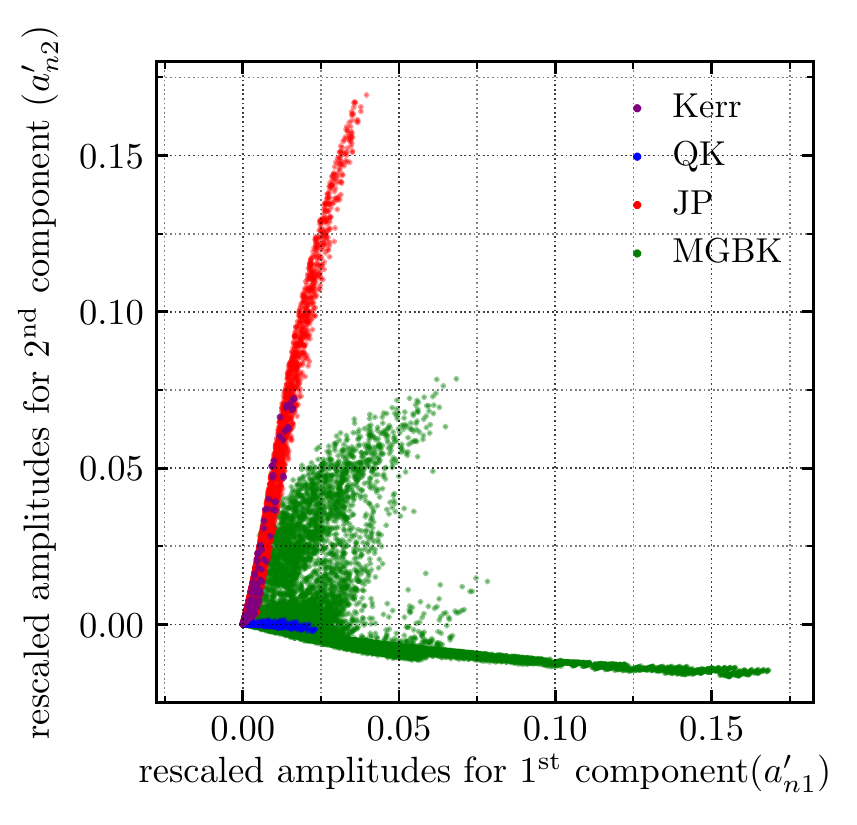}
\caption{The amplitude of the second PCA component for each shadow in our ensemble as a function of the amplitude of the first PCA component. The MGBK metric is shown in green, the JP metric in red, the QK metric in blue, and the Kerr metric in purple. Different metrics probe different regions of the parameter space spanned by the first two PCA components. }
\label{fig:colored_blobs}
\end{figure}

The accuracy with which metric-specific constraints can be imposed and the ability to break degeneracies
between different deviation parameters will depend on both the availability of prior information on the
properties of the observed black hole and on the interferometric coverage of the observations. For example,
as Figure~\ref{fig:params_JP} shows, in the JP metric, a nearly circular black-hole shadow can be obtained either for small
spin and minimal deviations from Kerr or large spin and a large value for the deviation parameter $\alpha_{13}$.
The two cases, however, result in substantially different shadow sizes, which will correspond to different
underlying black-hole masses. Prior knowledge of the mass of the black hole, e.g., from monitoring the
orbits of stars around it, as is the case for Sgr~A*, will help in breaking such a degeneracy. Similarly, nearly
circular shadows can also be obtained by reducing the inclination of the observer with respect to the spin
axis of the black hole. Prior information of this inclination, as is the case for M87, will be valuable in breaking
such degeneracies.

The current interferometric coverage available for EHT observations allows for measuring only global properties
of the black-hole shadows, such as their sizes and overall asymmetries, and not for tracing out their precise
shapes \citep{EHTPaperVI}. As Figures~\ref{fig:JP_x0_rscale} and~\ref{fig:MGBK_x0_rscale} show, such general measurements can be used primarily to place correlated constraints
on deviation parameters for different metrics and not to measure these parameters independently. The addition of
more EHT baselines on Earth and, in the future, a more complete interferometric coverage using rapidly orbiting,
space-based stations will allow measuring the precise shapes of black-hole shadows and reduce possible correlations
between parameters that describe deviations from the Kerr metric.

A set of measured PCA amplitudes of the shadow shape may also allow us to distinguish between different metrics. Figure \ref{fig:colored_blobs} shows the distribution of amplitudes of the first two PCA components for each metric we considered here. Measuring a large positive ratio between the 2nd and the 1st PCA amplitudes will point towards the Kerr metric or modifications similar to those described within the JP metric. On the other hand, measuring a negative ratio between the 2nd and the 1st PCA amplitudes will point towards modifications similar to those described in the MGBK metric.

The eigenshadows we derived can be directly applied to the results of EHT observations (see, e.g., \citealt{EHTPaperVI}) as they can be incorporated into a Hough/Radon transform and utilized in an edge detection algorithm (see~\citealt{2015ApJ...814..115P}).  The location of the shadow in the image plane and the size of the shadow ($R_{\mathrm{scale}}$) can also be incorporated into the Hough/Radon transform and compared to the $R_{\mathrm{scale}}$ and $x_0$ of the shadows in our data set. The outcome of this application will be a measurement of (or a constrain on) the coefficients of the various PCA components. This measurement will be agnostic of the metric of the compact object. 

\acknowledgements

\acknowledgements
We thank C. K. Chan, T. Lauer, Z. Younsi, C. Raithel, D. Ball, J. Kim, T. Trent, M. Rose, and M. de Laurentis for useful discussions. 
L.\;M.\ gratefully acknowledges support from NSF GRFP grant DGE~1144085, NSF PIRE 
grant 1743747, and an NSF Astronomy and Astrophysics Postdoctoral Fellowship under award no. AST-1903847.
D.\;P.\, and F.\;O.\, gratefully acknowledge support from NSF PIRE 
grant 1743747, NSF AST-1715061, and Chandra Award No. TM8-19008X 
for this work. All ray tracing and PCA calculations were performed with the \texttt{El~Gato}
GPU cluster at the University of Arizona that is funded by NSF award
1228509. 

\bibliography{main,my}
\end{document}